\documentclass{aastex63}
\usepackage{lineno}
\usepackage{graphicx}
\usepackage{gensymb}

\usepackage{graphicx}
\usepackage{wrapfig}
\usepackage{sidecap}
\usepackage{float}
\usepackage{array}
\usepackage{booktabs}
\usepackage{soul}
\usepackage{tablefootnote}
\usepackage{sidecap}
\usepackage{orcidlink}

	\shorttitle{JWST observations of Europa}
	\shortauthors{Cartwright et al.}
	
\begin{document}
            
            \title{JWST Reveals Spectral Tracers of Recent Surface Modification on Europa}

			\correspondingauthor{Richard J. Cartwright}
			\email{richard.cartwright@jhuapl.edu}
			
			\author{\orcidlink{0000-0002-6886-6009} Richard J. Cartwright$^a$}
			\affiliation{Johns Hopkins University Applied Physics Laboratory, 11101 Johns Hopkins Rd, Laurel, MD 20723}
			\footnote{$^a$Visiting Astronomer at the Infrared Telescope Facility, which is operated by the University of Hawaii under contract 80HQTR24DA010 with the National Aeronautics and Space Administration. }

            \author{\orcidlink{0000-0001-9089-4391} Charles A. Hibbitts}
			\affiliation{Johns Hopkins University Applied Physics Laboratory, 11101 Johns Hopkins Rd, Laurel, MD 20723}
   
       	    \author{\orcidlink{0000-0002-6117-0164} Bryan J. Holler}
			\affiliation{Space Telescope Science Institute, 3700 San Martin Drive, Baltimore, MD 21218} 
    		
            \author{\orcidlink{0000-0002-6036-1575} Ujjwal Raut}
			\affiliation{Southwest Research Institute, 6220 Culebra Road, San Antonio, TX 78238-5166}
              
            \author{\orcidlink{0000-0001-5888-4636} Tom A. Nordheim}
            \affiliation{Johns Hopkins University Applied Physics Laboratory, 11101 Johns Hopkins Rd, Laurel, MD 20723}                    
            \author{\orcidlink{0000-0002-6220-2869} Marc Neveu}
            \affiliation{University of Maryland, 4296 Stadium Dr, College Park, MD 20742}
            \affiliation{Solar System Exploration Division, NASA Goddard Space Flight Center, 8800 Greenbelt Road, Greenbelt, MD 20771}
            
            \author{\orcidlink{0000-0001-8541-8550} Silvia Protopapa}
            \affiliation{Southwest Research Institute, 1301 Walnut Street, Boulder, CO 80302}

            \author{\orcidlink{0000-0002-2161-4672} Christopher R. Glein}
			\affiliation{Southwest Research Institute, 6220 Culebra Road, San Antonio, TX 78238-5166}
            
            \author{\orcidlink{0000-0002-5150-5426} Erin J. Leonard}
            \affiliation{Jet Propulsion Laboratory, California Institute of Technology, 4800 Oak Grove Drive Pasadena, CA 91109} 
            
            \author{\orcidlink{0000-0003-0554-4691} Lorenz Roth}
            \affiliation{Space and Plasma Physics, KTH Royal Institute of Technology, Stockholm, Sweden}
            
            \author{\orcidlink{0000-0001-5048-6254} Chloe B. Beddingfield}
			\affiliation{Johns Hopkins University Applied Physics Laboratory, 11101 Johns Hopkins Rd, Laurel, MD 20723}
            
            \author{\orcidlink{0000-0002-2662-5776} Geronimo L. Villanueva}
			\affiliation{Solar System Exploration Division, NASA Goddard Space Flight Center, 8800 Greenbelt Road, Greenbelt, MD 20771}

			
			
	   \begin{abstract}
        Europa has been modified by a variety of geologic processes, exposing internally-derived materials that are heavily irradiated by charged particles trapped in Jupiter's magnetosphere. Prior spectral analysis of H$_2$O ice on Europa relied on low signal-to-noise data at wavelengths $>$2.5 $\micron$, limiting assessment of a 3.1 $\micron$ Fresnel peak that is diagnostic of exposed crystalline ice. We report new measurements of H$_2$O ice spectral features using high signal-to-noise data collected by the NIRSpec spectrograph (1.48 -- 5.35 $\micron$) on the James Webb Space Telescope. These data reveal a narrow 3.1 $\micron$ crystalline H$_2$O ice Fresnel peak, which is primarily located at southern latitudes in Tara and Powys Regiones. Our analysis indicates that crystalline ice exposed in these low-latitude regiones is likely sustained by ongoing thermal (re)crystallization outpacing charged particle amorphization of the top $\sim$10 $\micron$ of Europa's regolith over short timescales ($<$15 days). We also measured H$_2$O ice features centered near 1.5 $\micron$, 1.65 $\micron$, and 2.0 $\micron$, and a broad 3.6 $\micron$ H$_2$O continuum peak, which are all stronger at northern latitudes, in contrast to the 3.1 $\micron$ Fresnel peak identified at southern latitudes. These results support the hypothesis that H$_2$O ice in Europa's regolith is vertically stratified, with amorphous ice grains dominating its exposed surface, except in Tara and Powys Regiones. We also find that a previously detected 4.38 $\micron$ $^1$$^3$CO$_2$ feature is present almost exclusively at southern latitudes in Tara and Powys Regiones, likely derived from an internal source of carbon-bearing material.                
		\end{abstract}

		
		
		\section{Introduction} 
        Jupiter's moon Europa exhibits widespread evidence for recent endogenic activity \citep[]{greeley1998Europageol,greeley2000Europageol,greeley2004Europageol, pappalardo1999Europaocean, kattenhorn2014Europasubsump}, primarily resulting from tidal heating of its interior, driven by shared orbital resonances with the neighboring moons Io and Ganymede \citep[e.g.,][]{sotin2009Europatides}. This ongoing heating likely helps sustain an internal, global saline ocean, wedged between Europa's silicate-rich mantle and exterior icy shell \citep[e.g.,][]{anderson1998europa, petricca2025Europadifferentation}. Europa's induced magnetic field, first detected by the Galileo spacecraft, likely originates from interactions between its internal saline ocean and Jupiter's magnetosphere \citep{khurana1998Europainduced, kivelson2000Europainduced, zimmer2000Europainduced, schilling2007Europainduced}. Closer to the surface of Europa, pockets of saline meltwater could be perched in its icy shell, where collapse and refreezing may cause the formation of chaos terrains \citep[e.g.,][]{greeley2004Europageol, schmidt2011Europachaos}. Europa chaos exhibits a spectrum of morphologies \citep{leonard2022Europachaos} and a multitude of different mechanisms have been proposed to explain how they formed (summarized in \citealt{daubar2024Clippergeol}). The aforementioned liquid H$_2$O lens collapse and refreezing mechanism  \citep{schmidt2011Europachaos} may result in saline solutions rich in sodium and other components reaching Europa's surface and freezing in these chaos terrains, possibly exposing fresh, ocean-derived compounds \citep[e.g.,][]{vu2016Europasalts, thomas2017Europabrines, chivers2023Europabrines}. These ocean-derived materials are then chemically modified by Europa's intense radiation environment, driving radiolytic production of hydrogen peroxide (H$_2$O$_2$) and possibly sulfates, formed in part by implantation of S$^n$$^+$ ions originating from volcanic venting at Io, as well as other compounds \citep[e.g.,][]{carlson1999EuropaH2O2, carlson2002Europasulfate, brown2013Europasalts}. Other sources of exogenic material, in particular dust grains originating on Jupiter's irregular satellites \citep[]{bottke2013blackrain, chen2024Jdust}, potentially deliver ammoniated materials (NH-bearing), silicates, and carbon-rich species that overturn and mix with Europa's irradiated regolith \citep[]{cartwright2024CallistoJWST, sharkey2025JWSTirregulars}, and perhaps also mix with native sources of ammoniated species \citep{tosi2024GanymedeJuno}. Dust-mixed and irradiated salts and other components could then be delivered to Europa's subsurface via geologic conduits and (re)mix with native meltwater \citep{kattenhorn2014Europasubsump, hesse2022Europabrine, carnahan2022surface}. The exposure and subsequent return of modified components could be a continuous cycle that diversifies Europa's ocean chemistry \citep{zolotov2009Europachemistry}. The evidence for recent geologic activity and ongoing subsurface-surface cycling of components makes chaos terrains potentially useful windows into Europa's interior. Thus, measuring the spectral properties of chaos-dominated regions with remotely sensed data may provide insight into Europa's ocean chemistry and potential habitability \citep[e.g.,][]{hand2007Europaenergy, hand2009Europaastrobio}. 

        Reflectance spectra collected by ground-based telescopes, space telescopes, and spacecraft have revealed a wealth of information, linking Europa's surface constituents to its internal ocean and tenuous atmosphere \citep{calvin1995Gmoons, brown1996EuropaNa,brown2001EuropaK, carlson1996GmoonsNIMS, carlson1999EuropaH2O2, mccord1998NIMS, filacchione2019EuropaJuno, trumbo2019EuropaNaCl, trumbo2022EuropaNaCl}. Sodium chloride (NaCl) detected on Europa \citep{trumbo2022EuropaNaCl} likely originates in its interior, reaching the surface in chaos-dominated regions \citep[e.g.,][]{hand2009Europaastrobio,hesse2022Europabrine}, such as the large Tara and Powys Regiones that are located primarily over southern latitudes on Europa's leading and anti-Jovian hemispheres, respectively \citep[e.g,][]{leonard2024Europamap}. Exposed NaCl is irradiated by charged particles, forming optically active defects called color centers (absorption band near 0.46 $\micron$; \citealt{hand2015europa, trumbo2019EuropaNaCl, hibbitts2019Europasalts}). Continual irradiation gradually fragments and sputters NaCl molecules from Europa's surface \citep{brown2013Europasalts}, likely delivering Na to its exosphere \citep{brown1996EuropaNa} and perhaps to its neutral torus as well \citep{smith2019EuropaNatorus}. Similarly, the Cl$^+$ and Cl$^-$ ions detected in a pick up cloud proximal to Europa \citep[e.g.,][]{volwerk2001EuropaCl-} may originate from the destruction of NaCl on its surface. In contrast to the chaos-dominated regiones on Europa's leading and anti-Jovian hemispheres, the spectral signature of NaCl is less apparent on Europa's trailing hemisphere \citep{trumbo2019EuropaNaCl, trumbo2022EuropaNaCl}, possibly because of greater interactions with Jupiter's co-rotating plasma, which rapidly sputters and fragments NaCl, likely forming new salts via radiolysis, such as Na and Mg-bearing sulfates (MgSO$_4$), along with hydrated sulfuric acid (H$_2$SO$_4$ $\cdot$ XH$_2$O) \citep{carlson2002Europasulfate, brown2013Europasalts}.
        
       Along with salts, chaos terrains in Tara and Powys Regiones show spectral features indicative of other non-ice species, including a 3.51 $\micron$ H$_2$O$_2$ shoulder band, likely forming from irradiation of H$_2$O ice \citep{carlson1999EuropaH2O2, hand2013EuropaH2O2, trumbo2019EuropaH2O2, wu2024europaH2O2, raut2024EuropaH2O2}. Observations of Europa's leading and anti-Jovian hemispheres made by the Near-Infrared Spectrograph (NIRSpec) on the James Webb Space Telescope (JWST) revealed a double-lobed absorption feature, centered near 4.25 $\micron$ and 4.27 $\micron$, primarily concentrated in Tara and Powys Regiones, possibly representing ocean-derived CO$_2$ \citep{villanueva2023EuropaJWST, trumbo2023EuropaCO2}. Alternatively, this double-lobed feature may result from charged particle bombardment of carbonaceous material exposed in chaos terrains, forming CO$_2$ molecules through radiolysis \citep[e.g.,][]{gomis2005radiolyticCO2, raut2012radiolyticCO2, strazzulla2023radiolyticCO2}. A weak 4.38 $\micron$ absorption feature attributed to solid-state $^1$$^3$CO$_2$ was also detected in these JWST data \citep{villanueva2023EuropaJWST}. However, the spatial relationship between this 4.38 $\micron$ feature and the double-lobed 4.25 $\micron$ and 4.27 $\micron$ CO$_2$ band has yet to be determined, limiting our understanding of the origin and nature of CO$_2$ on Europa.
        
        The surface composition of Europa is primarily defined by H$_2$O, as ice \citep{pilcher1972GmoonsH2O}, as well as in hydrates \citep[e.g.,][]{mccord1998Europasalts}. The spectral signature of H$_2$O ice, in particular the fractional ratio between crystalline and amorphous ice, can be used to decipher the processes modifying icy regoliths and provide estimates for the rate of ice amorphization via charged particle bombardment. Prior analysis of data collected with Galileo's Near Infrared Mapping Spectrometer (NIMS; \citealt{carlson1992NIMS}) found that a 1.65 $\micron$ band, diagnostic of crystalline H$_2$O ice \citep[]{grundy1998temperatureH2O, mastrapa2008H2Oopcon}, is present across Europa's surface \citep{hansen2004Gmoonsice}. However, a prominent 3.1 $\micron$ Fresnel peak, also diagnostic of crystalline H$_2$O ice \citep{hagen1981H2Oice, mastrapa2009H2Oopcon}, is absent from Galileo/NIMS data of Europa, which instead exhibit a muted and somewhat neutral continuum in this wavelength range, more consistent with amorphous ice \citep{hansen2004Gmoonsice}. H$_2$O ice absorption is very strong in the wavelength range of the 3.1 $\micron$ Fresnel peak, causing photons to sample extremely shallow depths as they are effectively reflected off the surfaces of exposed ice grains \citep{hansen2004Gmoonsice}. The apparent disparity between these prior NIMS measurements could therefore result from a stratified regolith, with efficient amorphization of exposed H$_2$O ice by heavy ions and protons, probed by the 3.1 $\micron$ Fresnel peak ($<$1 $\micron$ depths; \citealt{mastrapa2009H2Oopcon}), whereas crystalline ice is  preserved beneath this topmost layer and is probed by the 1.65 $\micron$ band ($\sim$300 -- 500 $\micron$ depths; \citealt{mastrapa2008H2Oopcon}). Although analysis of Galileo/NIMS data advanced our understanding of H$_2$O ice on the Galilean moons, the generally low signal-to-noise (S/N) of these data, especially for Europa and other targets deep within the Jovian magnetosphere, has stymied follow-up work, in particular at wavelengths $>$2.5 $\micron$, where radiation-induced detector noise dominates and H$_2$O ice and hydrated salts absorb strongly, limiting the utility of older reflectance datasets. 
        
        Subsequent observations made by Juno's Jovian Infrared Auroral Mapper (JIRAM; \citealt{adriani2017JIRAM}) did reveal a 3.1 $\micron$ H$_2$O ice Fresnel peak, stronger at southern latitudes compared to northern latitudes on Europa's leading hemisphere \citep{filacchione2019EuropaJuno}, hinting that crystalline ice might be present at its exposed surface in some regions. However, JIRAM does not have sufficient wavelength coverage ($\sim$2 -- 5 $\micron$) to detect Europa's 1.65 $\micron$ band and cannot be used to simultaneously compare these diagnostic crystalline ice features. Although ground-based observations have mapped the distribution of Europa's 1.65 $\micron$ band \citep[e.g.,][]{ligier2016EuropaVLT}, strong absorption by H$_2$O ice on Europa's surface and persistent contamination by H$_2$O vapor and other gases in Earth's atmosphere have prevented reliable characterization and spatial mapping of the 3.1 $\micron$ Fresnel region. Thus, no single platform has been able to simultaneously measure and map Europa's 1.65 $\micron$ and 3.1 $\micron$ features in disk-resolved data since the Galileo mission ended in 2003, limiting our understanding of H$_2$O ice on Europa and the fractional abundance of amorphous and crystalline ice. 

        NIRSpec on JWST is ideal for simultaneously measuring and mapping the spatial distribution of the 1.65 $\micron$ and 3.1 $\micron$ crystalline H$_2$O ice features, as well as the 4.38 $\micron$ $^1$$^3$CO$_2$ feature. Here, we present new measurements and spectral maps for these three features, as well as H$_2$O ice features centered near 1.5, 2.0, and 3.6 $\micron$ using JWST/NIRSpec reflectance spectra (1.48 -- 5.35 $\micron$). We also present measurements and spectral maps of the 4.25 $\micron$ and 4.27 $\micron$ lobes of Europa's prominent solid-state CO$_2$ feature, which were reported in prior work \citep{villanueva2023EuropaJWST, trumbo2023EuropaCO2}. We use these measurements to investigate the processes modifying Europa's surface, including possible geologic exposure and subsequent irradiation of ocean-derived components and thermal recrystallization of H$_2$O ice in lower albedo regions like Tara and Powys. This study provides useful precursor context for upcoming Europa observations that will be made by Clipper's Mapping Imaging Spectrometer for Europa (MISE, 1 -- 5 $\micron$; \citealt{blaney2024ClipperMISE}), arriving at Jupiter in 2029 \citep[e.g.,][]{becker2024Clippercomp}. 

		\section{Data and Methods}

         \subsection{JWST/NIRSpec Reflectance Spectra} 
         
         \textit{Observation details.} As part of Guaranteed Time Observations (GTO) Program 1250, JWST observed Europa's leading hemisphere on November 23, 2022 (subobserver longitudes ranging between 91$\degree$ and 95$\degree$W, subobserver latitude 2.70$\degree$N). These observations were conducted with NIRSpec's Integral Field Unit (IFU) that has a 3$\arcsec$ x 3$\arcsec$ field of view and 0.1$\arcsec$ x 0.1$\arcsec$ spaxel dimensions across 30 image slices \citep{jakobsen2022NIRSpec, boker2023NIRSpec}. NIRSpec's G140H/F100LP, G235H/F170LP, and G395H/F290LP grating/filter combinations were each utilized during these observations, providing an average resolving power, R, of $\sim$ 2700 and nominal wavelength coverage between $\sim$1.1 to 5.3 $\micron$. Each of these H gratings record data across two detectors (short wavelength ``NRS1'' and long wavelength ``NRS2''), with ``unrecoverable'' wavelength gaps in between them where data are not recorded. These gaps shift slightly in wavelength space in each of the image slices that comprise NIRSpec's IFU. The Europa spectra exhibit wavelength gaps between 1.42 and 1.48 $\micron$ (G140H), 2.38 and 2.48 $\micron$ (G235H), and 4.02 and 4.18 $\micron$ (G395H). Each observation consisted of two exposures taken in different sections of the IFU (i.e., two dithers) for a total on-target time of $\sim$859 s with the G140H and $\sim$1074 s with the G235H and G395H, using the NRSRAPID readout mode.
         
         \textit{Saturation Mitigation.} Due to Europa's brightness at shorter wavelengths (1 -- 2.5 $\micron$), saturation occurred in nearly all the spaxels covering its disk in the G140H (NRS1 and NRS2) and G235H (NRS1) spectral cubes. To recover information in these spaxels, the {\it uncal} files were first trimmed from 4 groups down to 1 group, thereby eliminating the saturated groups in the G140H (NRS2) and most of the G235H (NRS1) spaxels. A modest range of wavelengths were additionally trimmed from the G235H (NRS1) data to remove residual saturation, creating a small wavelength gap (1.890 -- 1.898 $\micron$). No information was recovered from the G140H NRS1 data as saturation occurred within its first group, across all wavelengths (1.1 -- 1.42 $\micron$). In general, ``edge'' spaxels along the outer circumference of Europa's disk have lower signal-to-noise (S/N) and are more likely to retain artifacts after data processing. Thus, we focused our measurements and analyses on spaxels interior to this outer ring.

        \textit{Data Reduction.} The data were downloaded from the Mikulski Archive for Space Telescopes (10.17909/ve8q-sq65). Data processing was conducted with the Science Calibration Pipeline v1.14.0 with CRDS context jwst$\_$1241.pmap, processing raw {\it uncal} data into {\it s3d} spectral cubes for both dithers \citep{Bushouse2023JWSTpipe}. The pipeline was run with default parameters and the 1/$f$ pattern noise was removed using the NSClean routine \citep{rauscherNSClean}. The two dithers for each grating (NRS1 and NRS2) were georeferenced to Europa's disk and median combined, resulting in six total spectral cubes. To remove the solar component, we then divided each spaxel in each cube by a solar model generated with the Planetary Spectrum Generator (PSG; \citealt{villanueva2018PSG, villanueva2022PSG}). This PSG-derived solar model includes Doppler shifts, uses the ACE solar spectrum \citep[e.g.,][]{hase2010ace} to integrate Fraunhofer lines, and utilizes a well established solar model \citep{kurucz2005solar} to replicate the continuum intensity. Next, 1D spectra, and their associated georeferenced spatial information, were extracted from each spaxel on Europa's disk for all six spectral cubes. The 1D spectra were stitched together into one spectrum and then recombined into a finalized spectral cube, spanning 1.48 to 5.35 $\micron$ in each spaxel. Uncertainties for this spectral cube were estimated with standard error propagation routines that utilized the underlying calibrated uncertainties for each spaxel (as reported by the pipeline).    
        
\begin{figure}[h!]
   	\centering
   	\includegraphics[scale=0.85]{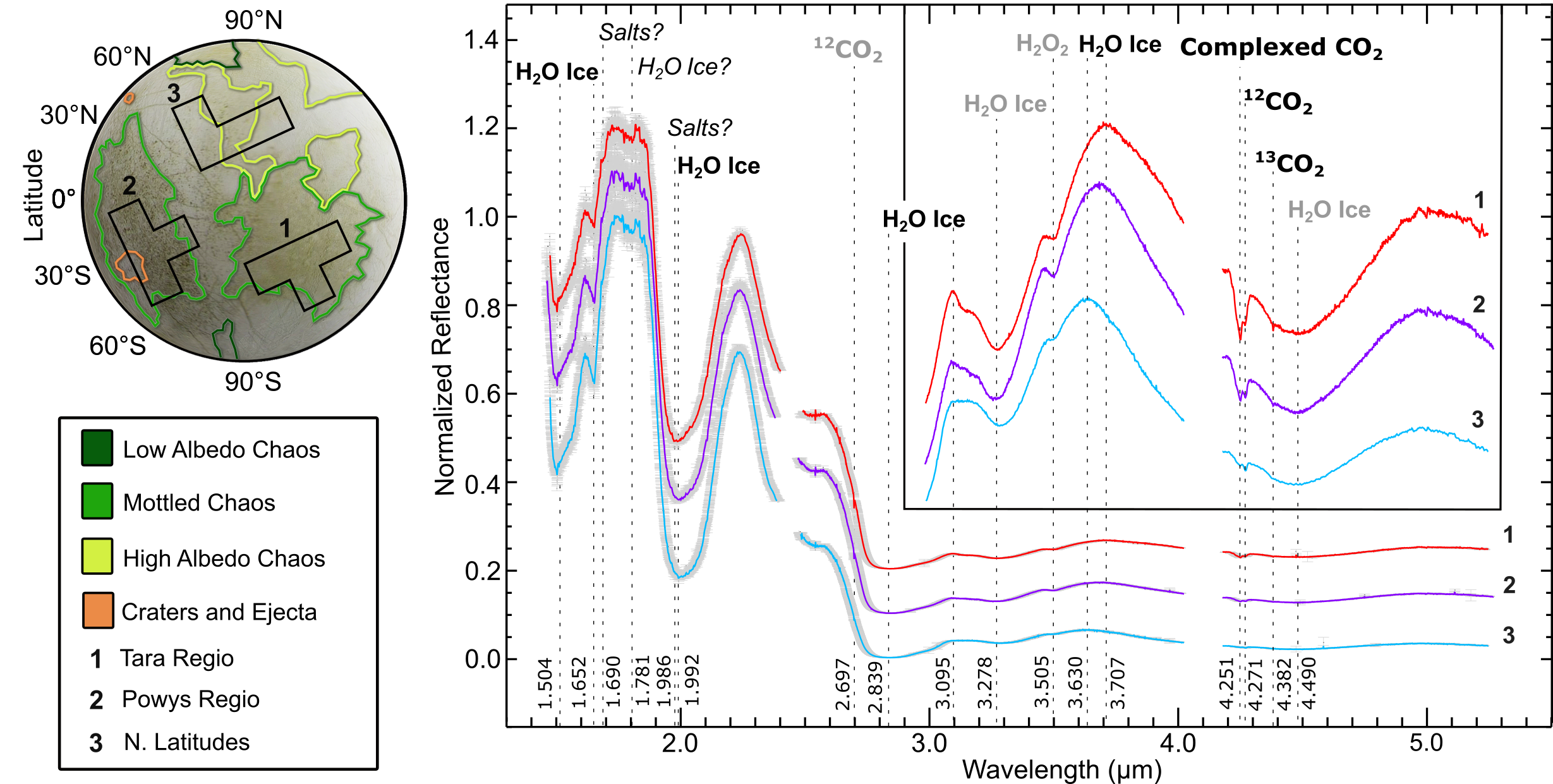}
   	\centering
   	\caption{\textit{Left: Mosaic of Europa's leading hemisphere, made with images collected by Galileo's Solid State Imager, overlain by mapping of previously defined geologic features \citep{leonard2024Europamap}. Black polygons show the locations of JWST/NIRSpec spaxels incorporated into the spectra shown on the right. Geologic units are defined in the legend below (utilizing a similar color scheme to \citealt{leonard2024Europamap}). Right: JWST/NIRSpec spectra and 1$\sigma$ uncertainties of Europa, representative of southern/central Tara Regio (red, 1), southern Powys Regio (purple, 2), and a northern low-latitude region dominated by ridges and bright terrains (cyan, 3). These spectra were generated by median-combining the four spaxels in each polygon shown in the map on the left, normalized to one between 1.77 and 1.78 $\micron$ and offset vertically for clarity. The inset figure shows the same data, between 3 and 5.3 $\micron$, normalized to one between 3.14 and 3.15 $\micron$ and offset vertically. Confirmed absorption bands associated with H$_2$O ice, solid-state CO$_2$, and H$_2$O$_2$, and the 3.1 $\micron$ H$_2$O ice Fresnel peak and 3.6 $\micron$ H$_2$O ice continuum peak are labeled (bolded text). Spectral features that might include contributions from hydrated minerals, such as sulfates \citep{deangelis2021Nasulfates}, chlorides \citep[e.g.,][]{deangelis2022MgCl}, and carbonates \citep[e.g.,][]{deangelis2019Nacarb}, are labeled ``salts?'' (italicized text). Central wavelengths ($\micron$) for these confirmed and possible features are listed vertically along each dotted line. Constituents labeled in gray font are shown here for completeness but were not measured by this study. Subtle spectral features between 4.7 and 5.1 $\micron$ could be artifacts, and we did not analyze them in this study.}}
   	\label{example_spectra}
   \end{figure}

        \textit{Regional Reflectance Spectra.} We generated representative spectra that highlight the spectral properties in three different regions on Europa's leading hemisphere: Tara Regio, Powys Regio, and Europa's ice-rich northern latitudes, normalized to 1 between 1.77 and 1.78 $\micron$. These spectra were generated by median combining 4 spaxels in each location (shown in Figure \ref{example_spectra}). These spaxels were chosen because they are representative of the distinct spectral properties of the three regions, while also being spatially contiguous and avoiding lower S/N edge spaxels.
        
        \textit{Comparison to Synthetic Spectra.} We compared the regional reflectance spectra to one-layer H$_2$O ice spectral models (100 $\micron$ diameters), generated using the real, \textit{n}, and imaginary, \textit{k}, indices of refraction (i.e., ``optical constants'') for crystalline H$_2$O ice (120 K) and amorphous ice (120 K) measured in the laboratory \citep{mastrapa2008H2Oopcon, mastrapa2009H2Oopcon}. These models were derived from Mie scattering theory \citep{bohren2008Mie} to calculate the single scattering albedo for each constituent, which were then passed to Hapke equations to calculate the geometric albedo (i.e., 0$\degree$ phase angle) as a function of wavelength \citep{hapke2012theory}. These Hapke-Mie models do not account for Europa's intrinsic photometric properties.   Nevertheless,  the models provide a useful approximation of H$_2$O ice measured in the laboratory that we can compare to H$_2$O ice features observed on Europa. Although Mie scattering can only approximate planetary regoliths, it has been widely applied to model the surfaces of icy bodies. Additional details and caveats on application of this Hybrid Hapke-Mie modeling approach for icy regoliths is provided in \citealt{cartwright2022ArielCO2, cartwright2024ArielJWST}.
        
		\subsection{Band Parameter Measurements} 
		
		We measured the areas and spectral contrast (i.e., band depths or peak heights) of all features using a measurement program that first defines and divides off a local linear or polynomial continuum for each feature \citep[e.g.,][]{cartwright2024CallistoJWST, cartwright2024ArielJWST}. Continuum-divided, spectral contrast measurements were made by averaging the reflectance values within $\pm$ 0.001 to 0.003 $\micron$ of the deepest/tallest point in each continuum-divided band (\textit{B$_c$}) or peak (\textit{P$_c$}) center, and uncertainties were computed using standard error propagation procedures \citep[e.g.,][]{taylor1997introduction}. The quality of these central wavelength positions are assessed manually before measuring the spectral contrast for each absorption feature (1 - \textit{B$_d$}) and scattering peak (\textit{P$_d$}  - 1). Using the trapezoidal rule, we measured each feature's continuum-divided area, with error estimation conducted with Monte Carlo-based sampling of the 1$\sigma$ errors for the spectral channels spanning each feature. 
        
        \textit{Band Measurement Caveats.} Because the G140H NRS1 data are saturated, the short wavelength end and associated continuum for Europa's 1.5 $\micron$ band are missing from these spectra, and we utilized a short wavelength cutoff of 1.48 $\micron$ for all measurements of this feature (Table \ref{band_measurements}). Consequently, the 1.5 $\micron$ band areas and spectral contrasts reported here are likely smaller than measurements made with other datasets that are able to sample the entire wavelength range of the 1.5 $\micron$ band. The 1.65 $\micron$ band (1.624 -- 1.682 $\micron$) is fully encapsulated within the broader 1.5 $\micron$ band (1.48 -- 1.742 $\micron$), and we therefore subtracted the 1.65 $\micron$ band area from the 1.5 $\micron$ band area to separate the contributions from each feature. Because of the long wavelength cutoff for the G395H NRS1 detector ($\sim$4 $\micron$), the long wavelength end of the 3.6 $\micron$ H$_2$O ice continuum peak was restricted to 4 $\micron$ for all measurements. The short wavelength end of the 3.6 $\micron$ H$_2$O ice continuum peak varies between $\sim$3.3 and 3.55 $\micron$, depending on its shape (Figure \ref{H2Obands_zoomin}) and central wavelength (Table \ref{band_measurements}). The 4.25 $\micron$ and 4.27 $\micron$ lobes of Europa's solid-state CO$_2$ feature are convolved, and we therefore report the same band area measurement for both lobes. Band minima for each lobe are readily discernible, and we are able to report separate band depths (Table \ref{band_measurements}). 

        \textit{Continuum-Divided Spectral Maps.} We generated spaxel maps using the central wavelength for each feature and their associated continuum-divided band area and spectral contrast measurements (shown in Figures \ref{H2O_maps1}, \ref{H2O_maps2}, and \ref{CO2_maps}). These maps are not projected and instead utilize the same grid of IFU spaxels that covered Europa's disk during the NIRSpec observations. Because we used continuum-divided measurements to make the maps, albedo variations between spaxels are effectively removed, and we therefore did not perform photometric corrections.
  
		\section{Results and Analyses} 
	 
	 	\subsection{Detected Spectral Features} 
	 	
\begin{figure}[h!]
	\centering
	\includegraphics[scale=0.90]{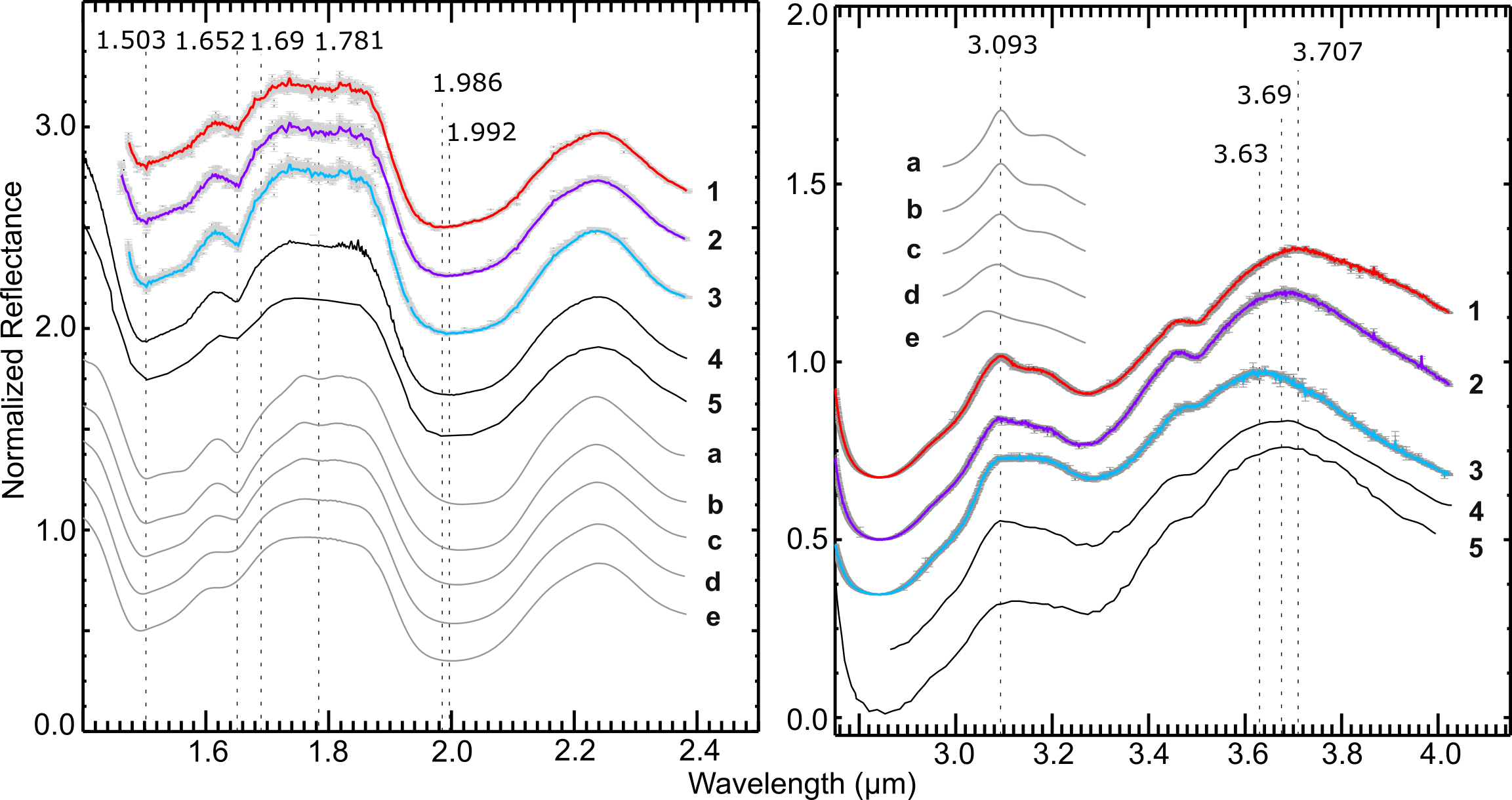}
	\centering
	\caption{\textit{Left: Tara (red, 1), Powys (purple, 2), and northern low-latitude (cyan, 3) JWST/NIRSpec spectra and 1$\sigma$ errors shown in Figure \ref{example_spectra} compared to a ground-based, disk-integrated IRTF/SpeX spectrum of Europa's leading hemisphere (black, 4, originally presented in \citealt{cartwright2023EuropaIRTF}) and a global spectrum collected by Galileo/NIMS (black, 5, originally reported in \citealt{mccord1998Europasalts}). These empirical data are compared to spectral models (gray) that utilize linear mixtures of 100 $\micron$ diameter grains of H$_2$O ice measured at 120 K \citep{mastrapa2008H2Oopcon, mastrapa2009H2Oopcon} with the following fractional abundances: (a) 100$\%$ crystalline H$_2$O ice, (b) 75$\%$ crystalline H$_2$O ice and 25$\%$ amorphous H$_2$O ice, (c) 50$\%$ crystalline H$_2$O ice and 50$\%$ amorphous H$_2$O ice, (d) 25$\%$ crystalline H$_2$O ice and 75$\%$ amorphous H$_2$O ice, and (e) 100$\%$ amorphous H$_2$O ice (see Section 2.1 for more detail on our modeling procedures). All spectra are normalized to 1 at 1.77 $\micron$ and offset vertically for clarity. The dotted lines show the central wavelengths of Europa's 1.5 $\micron$ and 1.65 $\micron$ H$_2$O ice bands, a weak 1.69 $\micron$ shoulder feature possibly resulting from salts, a potential hydrated salt or crystalline ice feature near 1.78 $\micron$, and the varying central wavelength for Europa's 2.0 $\micron$ H$_2$O ice band, shifting from 1.986 $\micron$ (Tara) to 1.992 $\micron$ (Powys and northern low-latitudes). Right: The same data and models, focused on longer wavelengths. The dotted lines show the central wavelengths of Europa's 3.1 $\micron$ H$_2$O Fresnel peak and the large wavelength shift in its 3.6 $\micron$ H$_2$O ice continuum peak, ranging from 3.63 $\micron$ in the northern low-latitudes spectrum, 3.69 $\micron$ in the Powys spectrum, and 3.71 $\micron$ in the Tara spectrum.}}
	\label{H2Obands_zoomin}
\end{figure}

        \textit{H$_2$O Ice and Hydrated Salts.} The JWST/NIRSpec data display prominent 1.5 $\micron$, 1.65 $\micron$, and 2.0 $\micron$ absorption bands, and 3.1 $\micron$ scattering peaks and 3.6 $\micron$ continuum peaks, all of which primarily result from H$_2$O ice \citep[e.g.,][]{grundy1998temperatureH2O, mastrapa2008H2Oopcon, mastrapa2009H2Oopcon}. In some locations, the broader 3.1 $\micron$ scattering peak (hereon referred to as the ``3.1 $\micron$ Fresnel region'') is overprinted by an additional narrow peak centered near 3.1 $\micron$, primarily in spaxels co-located with Tara and Powys Regiones (Figure \ref{example_spectra}). We also identify a subtle feature centered near 1.69 $\micron$ on the shoulder of the 1.5 $\micron$ feature (1.67 -- 1.71 $\micron$) that appears to deviate from the spectral signature of ``pure'' H$_2$O ice (Figure \ref{H2Obands_zoomin}), consistent with prior work that suggested non-ice species are contributing to the $\sim$1.7 $\micron$ wavelength range \citep{ligier2016EuropaVLT}. Although CH-bearing organic species can express absorption features near 1.7 $\micron$ \citep[e.g.,][]{quirico1997near,clark2009organics}, there are no obvious indicators for the stronger C-H stretching modes between $\sim$3.2 and 3.5 $\micron$ (Figures ~\ref{example_spectra} and ~\ref{H2Obands_zoomin}), hinting that salts might be more likely to contribute to this 1.69 $\micron$ shoulder feature. Another subtle feature near 1.78 $\micron$ might be consistent with crystalline H$_2$O ice or hydrated salts (Figure \ref{H2Obands_zoomin}).
        
        Aside from lower S/N edge spaxels, the 1.5 $\micron$ and 1.65 $\micron$ bands exhibit only minor variations in their band centers (1.502 -- 1.505 $\micron$ and 1.652 -- 1.653 $\micron$, respectively, Figure  \ref{H2Obands_zoomin}). Ground-based and Galileo/NIMS data show similar trends for Europa's 1.65 $\micron$ band, with widely varying band strengths but only small changes in its central wavelength position across Europa's leading and trailing hemispheres \citep{cartwright2023EuropaIRTF}. In contrast, the center of the 2.0 $\micron$ band exhibits wavelength shifts between the different regions of Europa's leading hemisphere (1.994 $\pm$ 0.022 $\micron$), with its central wavelength shifted to shorter wavelengths in spaxels associated with Tara Regio (1.98 -- 1.99 $\micron$), consistent with band measurements made by prior studies \citep[e.g.,][]{mccord1998Europasalts, ligier2016EuropaVLT}. 
        
        As defined in this study, Europa's 3.1 $\micron$ region spans $\sim$2.9 to 3.3 $\micron$ with a flat, plateau-shaped top, consistent with global NIMS spectra and  ground-based data collected over Europa's leading hemisphere (Figure \ref{H2Obands_zoomin}). The narrow 3.1 $\micron$ Fresnel peak that overprints the 3.1 $\micron$ region, primarily in spaxels spatially associated with Tara and Powys, exhibits minor variations in its central wavelength (3.096 $\pm$ 0.008 $\micron$). Europa's 3.6 $\micron$ continuum peak exhibits a large spread in central wavelengths (3.664 $\pm$ 0.06 $\micron$; Figure \ref{H2Obands_zoomin}), primarily shifted to shorter wavelengths at northern latitudes ($\sim$3.63 $\micron$), and longer wavelengths in Powys ($\sim$3.69 $\micron$) and Tara ($\sim$3.71 $\micron$), consistent with previous work that observed a 3.6 $\micron$ continuum peak shifted closer to 3.7 $\micron$ in Tara Regio  \citep{fischer2016EuropaKeck}. The shape of the 3.6 $\micron$ continuum peak varies widely across Europa, primarily because of the variable band strength of the 3.505 $\micron$ H$_2$O$_2$ feature that forms a prominent shoulder band on its short wavelength side (Figure \ref{example_spectra}). We present continuum-divided band measurement maps for H$_2$O ice in Figures \ref{H2O_maps1} and \ref{H2O_maps2}.

        \textit{Solid-State CO$_2$.} We identified absorption bands near 2.70 $\micron$, 4.25 $\micron$, 4.27 $\micron$, and 4.38 $\micron$ that are all associated with solid-state CO$_2$, as described in prior work \citep{villanueva2023EuropaJWST, trumbo2023EuropaCO2}. The 4.25 $\micron$ and 4.27 $\micron$ bands are two partially convolved lobes of Europa's strongest CO$_2$ feature. The central wavelength of the 4.25 $\micron$ lobe exhibits only minor variation across Europa's leading hemisphere (4.25 $\pm$ 0.002 $\micron$), whereas the 4.27 $\micron$ lobe exhibits a slightly larger spread in its central wavelength (4.27 $\pm$ 0.005 $\micron$). The central wavelength of the 4.38 $\micron$ band is essentially locked to 4.381 or 4.382 $\micron$, and we detect no evidence for a double-lobed structure. Because we are primarily interested in understanding the relationship between Europa's 4.38 $\micron$ band and its double-lobed 4.25 $\micron$ and 4.27 $\micron$ CO$_2$ feature, we did not analyze the 2.70 $\micron$ feature here and instead refer the reader to prior work that investigated this band \citep{villanueva2023EuropaJWST, trumbo2023EuropaCO2}. We present continuum-divided band measurement maps for solid-state CO$_2$ in Figure \ref{CO2_maps}.
   
     \begin{table}[t]
	\caption {Band measurements for the regional spectra of Tara and Powys Regiones and the Northern Low-Latitudes Zone} \label{band_measurements} 
	\hskip-2cm\begin{tabular}{cccccccc}
		\hline\hline
		\begin{tabular}[c]{@{}l@{}}  \hspace{-1 cm}Region \end{tabular} &   
		\begin{tabular}[c]{@{}l@{}}\hspace{-1 cm}Feature \\ \hspace{-1 cm}Name \end{tabular} & 
		\begin{tabular}[c]{@{}l@{}}  \hspace{-1 cm}Feature \\ \hspace{-1 cm}Center \\ \hspace{-1 cm}($\micron$) \end{tabular} & 
		\begin{tabular}[c]{@{}l@{}}  \hspace{-1 cm}Feature \\ \hspace{-1 cm}Wavelength \\ \hspace{-1 cm} Range ($\micron$) \end{tabular} &   
		\begin{tabular}[c]{@{}l@{}}  \hspace{-1 cm}Spectral \\ \hspace{-1 cm}Contrast ($\%$) \end{tabular} &
		\begin{tabular}[c]{@{}l@{}}  \hspace{-1 cm}Band Area \\ \hspace{-1 cm}(10$^-$$^3$ $\micron$) \end{tabular} &		
		\begin{tabular}[c]{@{}l@{}}  \hspace{-1 cm}$>$3$\sigma$ Spectral \\ \hspace{-1 cm}Contrast and \\ \hspace{-1 cm}Band Area? \end{tabular} &
		\begin{tabular}[c]{@{}l@{}} \hspace{-1 cm}Constituents \\ \hspace{-1 cm}(confirmed, bolded) \\ \hspace{-1 cm}(suggested, italicized) \end{tabular} \\
		\hline
		\vspace{-0. cm}1. Tara & $^a$$^,$$^b$$^,$$^c$1.5 $\micron$ Band & 1.504 & 1.480 -- 1.742 & 39.62 $\pm$ 0.04 & 54.14 $\pm$ 0.56 & Yes & \textbf{Crystalline H$_2$O Ice,} \\
          2. Powys & & 1.502 & & 46.93 $\pm$ 0.33 & 71.29 $\pm$ 0.04 & Yes & \textit{Hydrated Salts}  \\
          3. N. Lat. & & 1.504 & & 56.93 $\pm$ 0.39 & 89.34 $\pm$ 0.06 & Yes & \\
        \hline
		\vspace{-0. cm}1. Tara & $^b$1.65 $\micron$ Band & 1.652 & 1.624 -- 1.682 & 9.58 $\pm$ 0.57 & 2.84 $\pm$ 0.03 & Yes & \textbf{Crystalline H$_2$O Ice} \\
          2. Powys & & 1.653 & & 13.00 $\pm$ 0.55 & 3.93 $\pm$ 0.02 & Yes & \\
          3. N. Lat. & & 1.652 & & 17.24 $\pm$ 0.64 & 5.13 $\pm$ 0.04 & Yes & \\
        \hline          
		\vspace{-0. cm}1. Tara & $^c$2.0 $\micron$ Band & 1.986 & 1.857 -- 2.237 & 67.06 $\pm$ 0.19 & 150.50 $\pm$ 0.54 & Yes & \textbf{Crystalline H$_2$O Ice,} \\
          2. Powys & & 1.992 & & 70.03 $\pm$ 0.34 & 156.86 $\pm$ 0.13 & Yes & \textit{Amorphous H$_2$O Ice,} \\
          3. N. Lat. & & 1.992 & & 78.31 $\pm$ 0.36 & 180.62 $\pm$ 0.07 & Yes & \textit{Hydrated Salts} \\ 
        \hline          
		 \vspace{-0. cm}1. Tara & 3.1 $\micron$ Region & 3.095 & 3.034 -- 3.280 & 37.87 $\pm$ 0.81 & 47.65 $\pm$ 0.30 & Yes & \textbf{Crystalline H$_2$O Ice,} \\
          2. Powys & & 3.091 & 3.045 -- 3.255 & 24.85 $\pm$ 0.49 & 30.85 $\pm$ 0.17 & Yes & \textbf{Amorphous H$_2$O Ice} \\
          3. N. Lat. & & 3.104 & 3.055 -- 3.289 & 15.26 $\pm$ 0.59 & 24.23 $\pm$ 0.18 & Yes & \\
        \hline          
		\vspace{-0. cm}1. Tara & 3.1 $\micron$ Peak & 3.093 & 3.060 -- 3.125 & 6.83 $\pm$ 0.96 & 2.08 $\pm$ 0.13 & Yes & \textbf{Crystalline H$_2$O Ice,} \\
          2. Powys & & 3.087 & & 3.36 $\pm$ 0.69 & 0.88 $\pm$ 0.07 & Yes & \textbf{Amorphous H$_2$O Ice} \\
          3. N. Lat. & & N/A & & -0.08 $\pm$ 1.15 & -0.06 $\pm$ 0.06 & No & \\
        \hline          
		\vspace{-0. cm}1. Tara & $^c$3.6 $\micron$ Peak & 3.707 & 3.537 -- 4.000 & 29.95 $\pm$ 0.12 & 479.98 $\pm$ 0.15 & Yes & \textbf{Crystalline H$_2$O Ice,} \\
          2. Powys & & 3.690 & 3.408 -- 4.000 & 47.07 $\pm$ 0.17 & 657.93 $\pm$ 0.26 & Yes & \textit{Hydrated Salts} \\
          3. N. Lat. & & 3.631 & 3.342 -- 4.000 & 68.68 $\pm$ 0.21 & 814.46 $\pm$ 0.31 & Yes & \\ 
        \hline          
		\vspace{-0. cm}1. Tara & $^d$4.25 $\micron$ Lobe & 4.251 & 4.203 -- 4.292 & 25.38 $\pm$ 0.35 & 10.36 $\pm$ 0.03 & Yes & \textbf{Complexed CO$_2$} \\
          2. Powys & & 4.251 & & 16.51 $\pm$ 0.34 & 7.01 $\pm$ 0.04 & Yes & \\
          3. N. Lat. & & 4.250 & & 6.52 $\pm$ 0.28 & 2.99 $\pm$ 0.05 & Yes & \\
        \hline          
		\vspace{-0. cm}1. Tara & $^d$4.27 $\micron$ Lobe & 4.269 & 4.203 -- 4.292 & 16.70 $\pm$ 0.23 & 10.36 $\pm$ 0.03 & Yes & \textbf{$^1$$^2$CO$_2$} \\
          2. Powys & & 4.271 & & 12.44 $\pm$ 0.30 & 7.01 $\pm$ 0.04 & Yes & \\
          3. N. Lat. & & 4.271 & & 7.87 $\pm$ 0.83 & 2.99 $\pm$ 0.05 & Yes & \\
        \hline          
		\vspace{-0. cm}1. Tara & 4.38 $\micron$ Band & 4.382 & 4.370 -- 4.388 & 4.06 $\pm$ 0.68 & 0.25 $\pm$ 0.02 & Yes & \textbf{$^1$$^3$CO$_2$} \\
          2. Powys & & 4.381 & & 2.54 $\pm$ 0.33 & 0.17 $\pm$ 0.01 & Yes & \\
          3. N. Lat. & & N/A & & 0.54 $\pm$ 0.39 & 0.09 $\pm$ 0.02 & No & \\            
		\hline
	\end{tabular}
 
    \vspace{0.1 cm}
    \hspace{0.05 cm} $^a$\textit{The 1.50 $\micron$ band is bound by the short wavelength cutoff of the G140H grating (NRS2) and the values reported here are likely underestimates.} \\
    $^b$\textit{The 1.65 $\micron$ feature fully overlaps the 1.5 $\micron$ feature, and thus, its band area was subtracted off the 1.5 $\micron$ band area.} \\
    $^c$\textit{These features may include contributions from ``hydrated salts,'' a catchall designation that includes NaCl.} \\
    $^d$\textit{Band areas for the 4.25 and 4.27 $\micron$ lobes are convolved, and we report the same band area for both features.} \\
    \end{table}

 \vspace{0.5cm}

	\subsection{Band Parameter Measurements} 
	
\begin{figure}[h!]
         \center
			\includegraphics[scale=0.75]{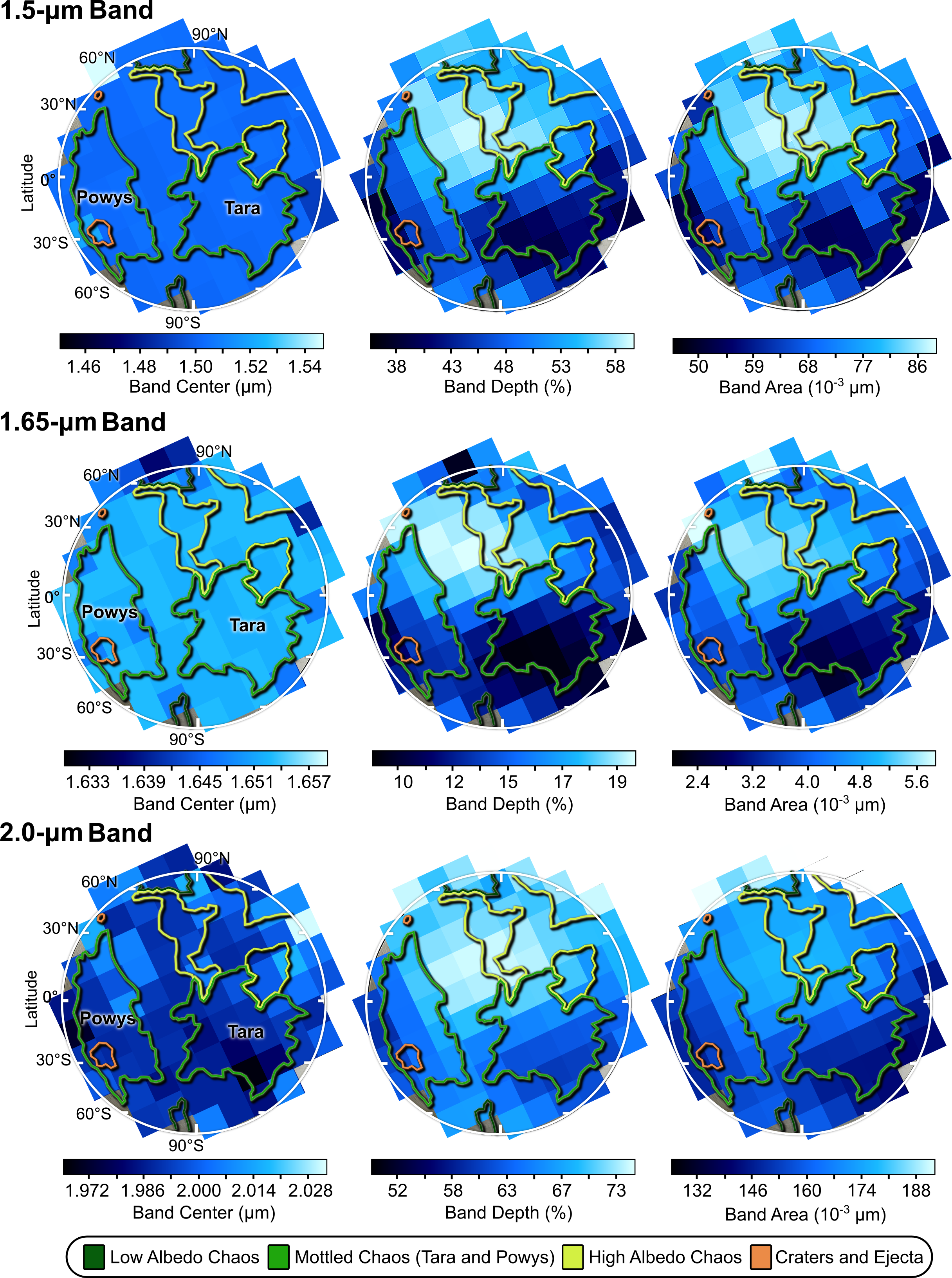}
		 \center 
            \vspace{-0.3 cm}\caption{\textit{Spectral maps illustrating the distribution in the band centers, depths, and areas for (top) Europa's 1.5 $\micron$ band (1.48 -- 1.74 $\micron$), (middle) its 1.65 $\micron$ band (1.62 -- 1.68 $\micron$), and (bottom) its 2.0 $\micron$ band (1.86 -- 2.24 $\micron$), each primarily attributed to H$_2$O ice. The 1$\sigma$ uncertainties for the spectral contrast measurements range between: 0.37 -- 0.57$\%$, 0.75 -- 1.11$\%$, and 0.10 -- 1.00$\%$, for the 1.5 $\micron$, 1.65 $\micron$, and 2.0 $\micron$ bands, respectively. The 1$\sigma$ uncertainties for the area measurements are 0.12 -- 0.32 $\micron$, 0.11 -- 0.21 $\micron$, and 0.08 -- 0.88 $\micron$ for the 1.5 $\micron$, 1.65 $\micron$, and 2.0 $\micron$ bands, respectively. The spatial extents of large-scale geologic units on Europa's leading and anti-Jovian sides, including Tara (10$\degree$S, 75$\degree$W) and Powys (0$\degree$, 145$\degree$W) Regiones are indicated and described in the included legend (see \citealt{leonard2024Europamap}).}}
        \label{H2O_maps1}
   \end{figure}

\begin{figure}[h!]
         \center
			\includegraphics[scale=0.75]{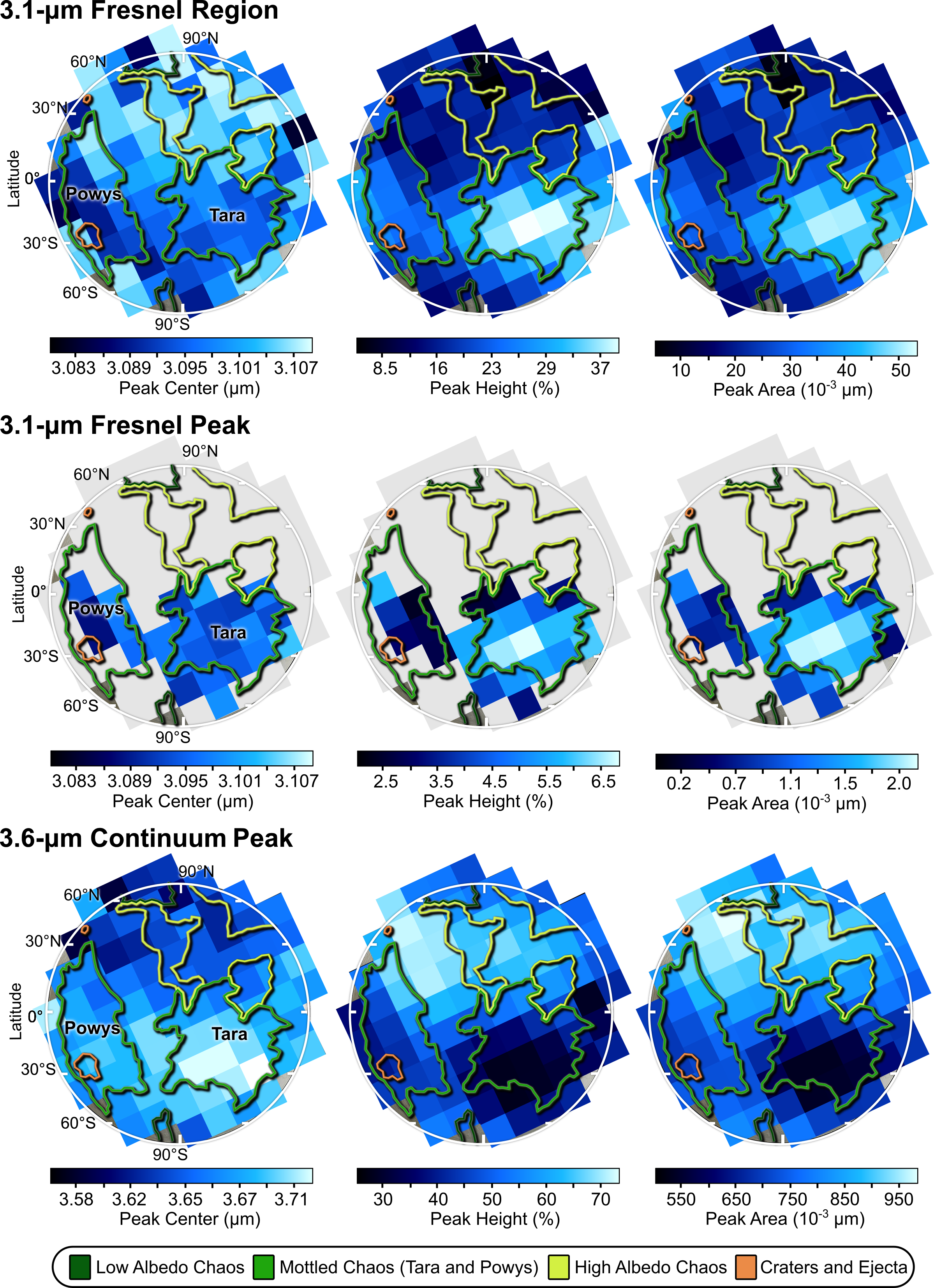}
		 \center
            \vspace{-0.3 cm}\caption{\textit{Spectral maps illustrating the distribution in the centers, heights, and areas for (top) Europa's 3.1 $\micron$ H$_2$O ice Fresnel region (2.9 -- 3.3 $\micron$), (middle) its 3.1 $\micron$ H$_2$O ice Fresnel peak (3.05 -- 3.15 $\micron$), and (bottom) its 3.6 $\micron$ H$_2$O ice continuum peak (3.3 -- 4.0 $\micron$). The 1$\sigma$ uncertainties for the spectral contrast measurements range between: 0.58 - 1.78$\%$, 0.70 -- 1.00$\%$, and 0.011 -- 0.79$\%$, for the 3.1 $\micron$ Fresnel region, 3.1 $\micron$ Fresnel peak, and 3.6 $\micron$ continuum peak, respectively. The 1$\sigma$ uncertainties for the area measurements are 0.080 - 0.094 $\micron$, 0.08 -- 0.26 $\micron$, and 0.08 -- 0.14 $\micron$ for Europa's Fresnel region, 3.1 $\micron$ Fresnel peak, and the 3.6 $\micron$ continuum peak, respectively. Spaxels where the 3.1 $\micron$ Fresnel peak was not detected ($>$3$\sigma$) are shown in light gray. The spatial extents of large-scale geologic units on Europa's leading and anti-Jovian sides, including Tara (10$\degree$S, 75$\degree$W) and Powys (0$\degree$, 145$\degree$W) Regiones are indicated and described in the included legend (see \citealt{leonard2024Europamap}).}}
        \label{H2O_maps2}
   \end{figure}

	     \textit{H$_2$O Ice and Hydrated Minerals.} We report spectral maps showing continuum-divided band areas and spectral contrast measurements (i.e., band depths and peak heights) for Europa's 1.5 $\micron$, 1.65 $\micron$, and 2.0 $\micron$ H$_2$O ice bands, and its 3.1 $\micron$ Fresnel region and peak and 3.6 $\micron$ H$_2$O peak (Figures \ref{H2O_maps1} and \ref{H2O_maps2}). The 1.5 $\micron$, 1.65 $\micron$, and 2.0 $\micron$ bands and 3.6 $\micron$ continuum peak show similar spatial trends, with larger spectral contrasts and areas at northern latitudes, particularly in a wedge-shaped region, east of Powys and northwest of Tara (represented by spectrum 3 in Figure \ref{example_spectra}), which was previously identified as an ice-rich region on Europa \citep[e.g.,][]{ligier2016EuropaVLT}. The 3.1 $\micron$ Fresnel region exhibits larger spectral contrasts and areas in spaxels associated with Tara and Powys Regiones over Europa's southern latitudes compared to its northern latitudes (Figures  \ref{H2O_maps2} and  \ref{3.10/1.65_ratio}), consistent with prior analysis of Juno/JIRAM data \citep{filacchione2019EuropaJuno}. Furthermore, the 3.1 $\micron$ Fresnel peak is only reliably observed at southern latitudes ($>$3$\sigma$ detection), with the strongest peaks measured in spaxels covering Tara Regio. We also conducted band parameter analyses on the Tara, Powys, and northern low-latitudes regional spectra (Table \ref{band_measurements}), finding the same trends in band area and spectral contrast measurements. Thus, our band parameter measurements confirm a north/south latitudinal dichotomy in the strength of H$_2$O ice features on Europa's leading hemisphere that was described in prior work \citep{ligier2016EuropaVLT, filacchione2019EuropaJuno}.
   
\begin{figure}[h!]
         \center
			\includegraphics[scale=0.75]{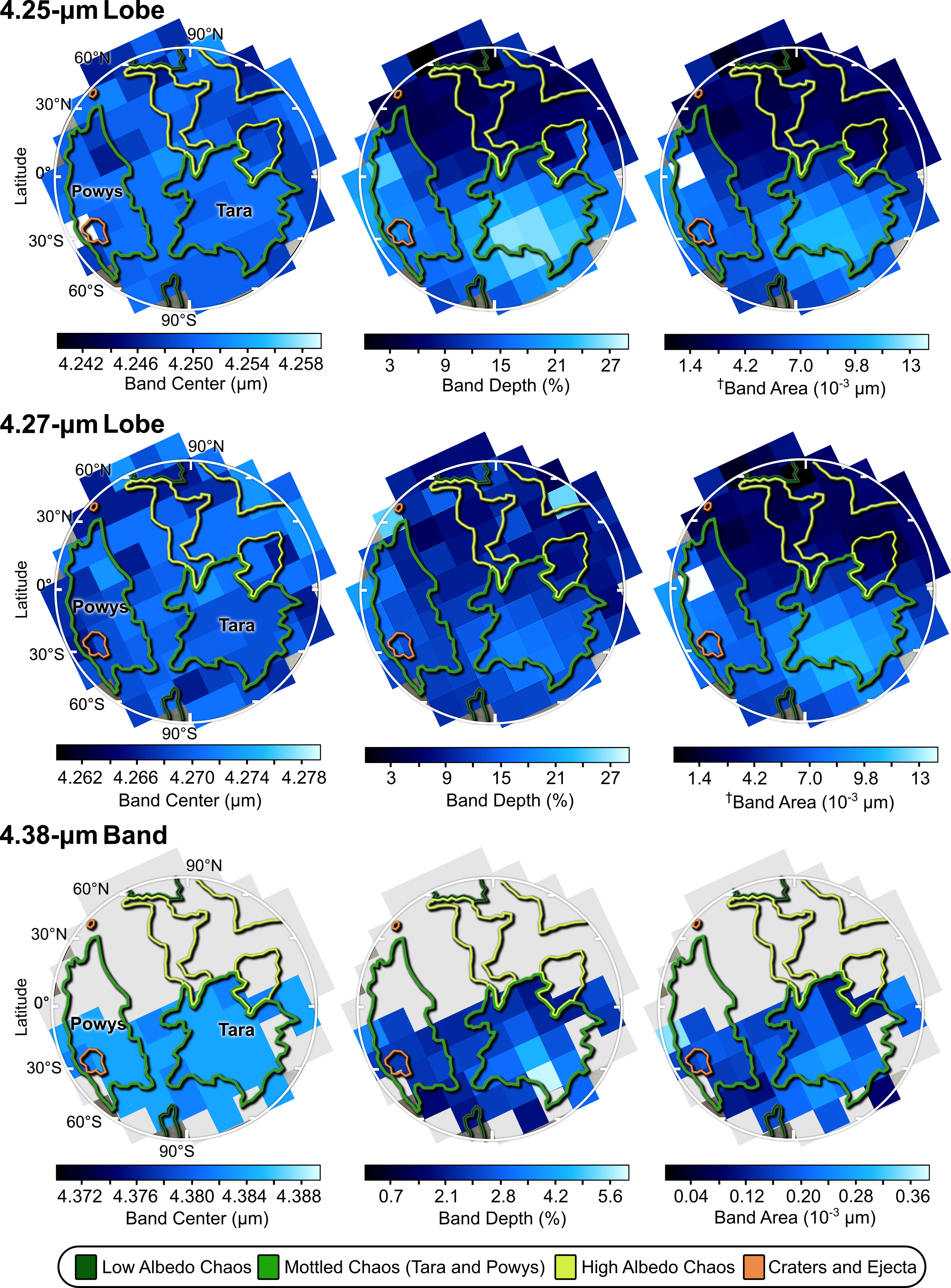}
		 \center
            \vspace{-0.3 cm}\caption{\textit{Spectral maps illustrating the distribution in the band centers, depths, and areas for (top) Europa's 4.25 $\micron$ lobe (4.2 -- 4.3 $\micron$), (middle) 4.27 $\micron$ lobe (4.2 -- 4.3 $\micron$), and (bottom) 4.38 $\micron$ band (4.37 -- 4.39 $\micron$), each attributed to solid-state CO$_2$. $^{\dagger}$The band areas of the 4.25 $\micron$ and 4.27 $\micron$ lobes are convolved, and we therefore report the same area measurements for both features. The 1$\sigma$ uncertainties for the spectral contrast measurements range between: 0.10 -- 0.90$\%$, 0.38 -- 1.90$\%$, and 0.40 -- 1.36$\%$, for the 4.25 $\micron$ region, 4.27 $\micron$ band, and 4.38 $\micron$ band, respectively. The 1$\sigma$ uncertainties for the area measurements are 0.03 -- 0.07 $\micron$ and 0.01 -- 0.03 $\micron$ for the convolved 4.25 $\micron$ and 4.27 $\micron$ lobes and the 4.38 $\micron$ band, respectively. Spaxels where the 4.38 $\micron$ band was not detected ($>$3$\sigma$) are shown in light gray. The spatial extents of large-scale geologic units on Europa's leading and anti-Jovian sides, including Tara (10$\degree$S, 75$\degree$W) and Powys (0$\degree$, 145$\degree$W) Regiones are indicated and described in the included legend (see \citealt{leonard2024Europamap}).}}
        \label{CO2_maps}
   \end{figure}

        \textit{Solid-state CO$_2$.} We report band depth and area measurements for Europa's double-lobed, solid-state CO$_2$ feature and its 4.38 $\micron$ $^1$$^3$CO$_2$ band, using spectral maps spanning Europa's leading hemisphere (Figure \ref{CO2_maps}). Our results indicate that the 4.25 $\micron$ and 4.27 $\micron$ lobes have stronger band depths at southern latitudes, primarily in spaxels associated with Tara and Powys, consistent with previously reported measurements \citep{villanueva2023EuropaJWST, trumbo2023EuropaCO2}. Additionally, the 4.25 $\micron$ lobe's band depths weaken considerably at northern latitudes, whereas the 4.27 $\micron$ lobe's band depths decrease more gradually at northern latitudes (Figure \ref{CO2_compare}). The convolved band areas for these two lobes show similar trends, with larger area measurements at southern latitudes, in particular in spaxels spanning Tara Regio, and weaker measurements at northern latitudes. The band depth and area measurements for the 4.38 $\micron$ band indicate that this feature is primarily observed at southern latitudes, spatially associated with Tara and Powys ($>$3$\sigma$ detection), and is essentially undetected at northern latitudes (Figure \ref{CO2_maps}). 
        
        Of note, a 4.38 $\micron$ feature detected on Ganymede with JWST/NIRSpec (G395H grating) was attributed to a data calibration artifact \citep{bockelee2024GanymedeJWST}. Europa's 4.38 $\micron$ feature is much narrower than the broad 4.38 $\micron$ feature detected on Ganymede, and it exhibits discernible band depth and area variations across Europa's disk that appear to track the spatial distribution of its double-lobed CO$_2$ feature (Figures \ref{CO2_maps} and \ref{CO2_compare}), unlike Ganymede's 4.38 $\micron$ feature that exhibits a mostly uniform profile across its surface. Thus, the measurements we report here demonstrate that the 4.38 $\micron$ feature very likely results from solid-state $^1$$^3$CO$_2$ detected on Europa's surface. 
      
   \section{Discussion}
   
    \subsection{Spatial Distribution and Vertical Stratification of H$_2$O Ice} 
    
\begin{figure}[h!]
         \center
			\includegraphics[scale=0.77]{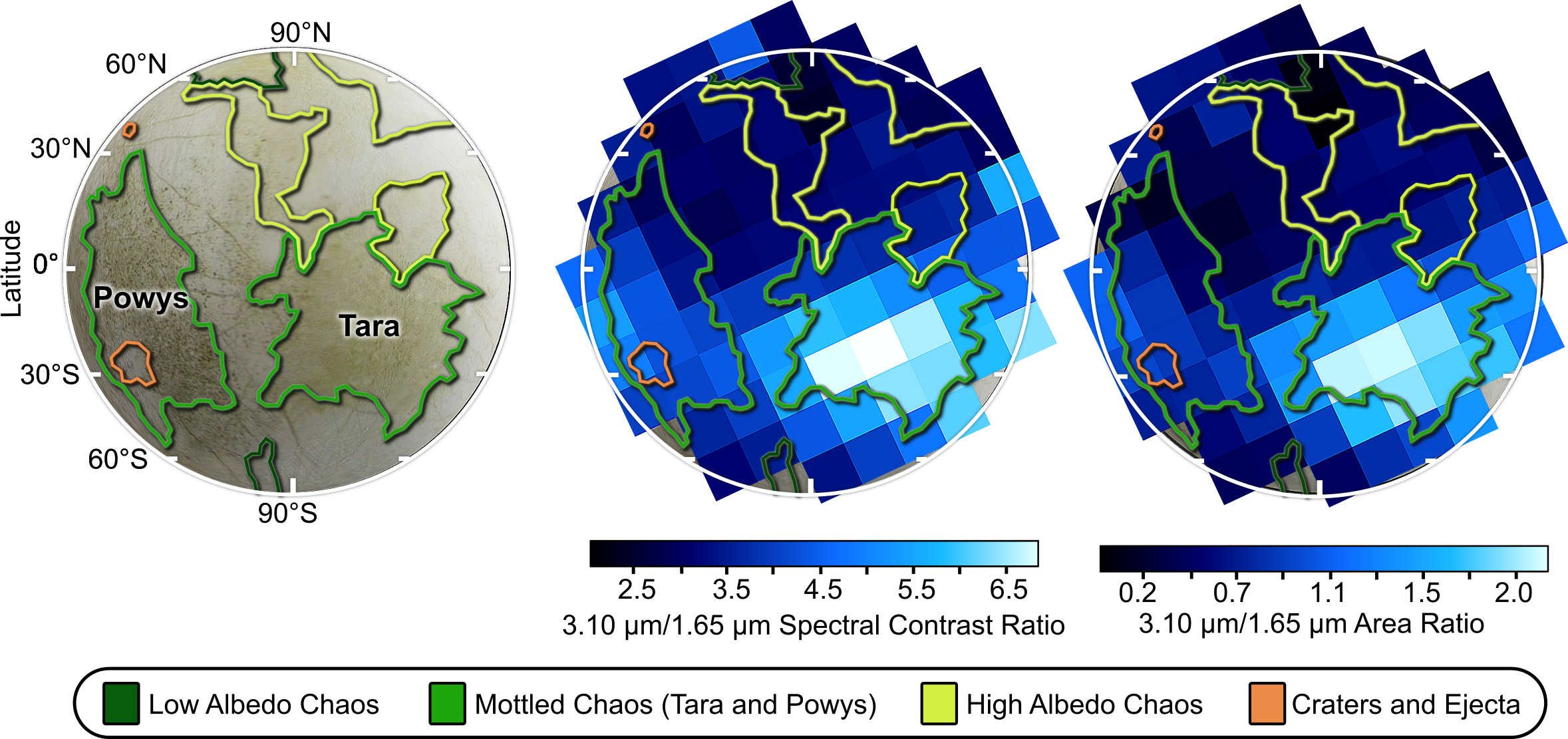}
		 \center
            \vspace{-0.1 cm}\caption{\textit{Spaxel maps displaying the ratios between Europa's 3.1 $\micron$ Fresnel region and its 1.65 $\micron$ band (3.1 $\micron$ Region/1.65 $\micron$ band), generated with the spectral contrast and area measurements shown in Figures \ref{H2O_maps1} and \ref{H2O_maps2}. These ratio maps highlight the dissimilar trends in the distribution of crystalline H$_2$O ice on Europa, with the 1.65 $\micron$ band stronger at northern latitudes and the 3.1 $\micron$ Fresnel region stronger at southern latitudes, particularly in Tara and Powys Regiones. This dissimilarity points to vertical stratification of H$_2$O ice in Europa's regolith, with more exposed crystalline H$_2$O ice at southern latitudes. The spatial extents of large-scale geologic units on Europa's leading and anti-Jovian sides, including Tara (10$\degree$S, 75$\degree$W) and Powys (0$\degree$, 145$\degree$W) Regiones are indicated and described in the included legend (see \citealt{leonard2024Europamap}).}}
        \label{3.10/1.65_ratio}
   \end{figure}

	    All of the H$_2$O features we measured are stronger at northern latitudes, except for the 3.1 $\micron$ H$_2$O ice Fresnel region and the narrow 3.1 $\micron$ Fresnel peak that are notably stronger at southern latitudes, in particular in spaxels spanning Tara and Powys Regiones (Figures \ref{H2O_maps1} and \ref{H2O_maps2}). This latitudinal dichotomy in the strength of Europa's H$_2$O ice features is particularly puzzling for the 1.65 $\micron$ band and the 3.1 $\micron$ Fresnel region (Figure \ref{3.10/1.65_ratio}), which are both diagnostic of crystalline ice \citep[e.g.][]{mastrapa2008H2Oopcon, mastrapa2009H2Oopcon}. One way to explain this discrepancy is if Europa's icy regolith is vertically stratified, and these H$_2$O ice features are sampling different depths, as suggested by prior work \citep{hansen2004Gmoonsice}. Laboratory experiments have demonstrated that photon penetration depths into H$_2$O ice are a steep function of wavelength, with photons spanning the 1.5 $\micron$, 1.65 $\micron$, and 2 $\micron$ bands traversing $\sim$100 to 500 $\micron$ depths into icy surfaces \citep[e.g.,][]{mastrapa2008H2Oopcon, mastrapa2009H2Oopcon}. In contrast, across the wavelength range of the H$_2$O ice Fresnel region ($\sim$2.9 -- 3.3 $\micron$), photons are largely absorbed within the very top surface of icy regoliths ($<$10 $\micron$ depths), and in the narrower wavelength range of the 3.1 $\micron$ Fresnel peak ($\sim$3.05 -- 3.15 $\micron$), photons essentially only interact with the surfaces of exposed H$_2$O ice grains ($<$1 $\micron$ depths; \citealt{becker2024Clippercomp}). H$_2$O ice features at shorter near-infrared wavelengths ($<$2.5 $\micron$) therefore probe greater depths than H$_2$O ice features in the Fresnel region, and comparing spectral features in these two wavelength ranges can provide key clues on the vertical stratification of H$_2$O-dominated regoliths \citep[e.g.,][]{hansen2004Gmoonsice, cartwright2015UmoonCO2, cartwright2020UmoonIRAC, protopapa2024CharonJWST}. 
	    
	     Prior work that investigated vertical stratification of Europa's icy regolith determined that its exposed surface is dominated by amorphous ice due to the non-detection of a narrow 3.1 $\micron$ Fresnel peak \citep{hansen2004Gmoonsice}. Our results show that the narrow 3.1 $\micron$ Fresnel peak is present on Europa's leading hemisphere, but only at southern latitudes, primarily in spaxels spanning Tara and Powys Regiones (discussed at greater length in Section 4.4). The much lower S/N of Galileo/NIMS data compared to JWST/NIRSpec data at wavelengths $>$2.5 $\micron$ might explain the prior non-detection of this weak feature.              
	     
   \subsection{Spectral Properties and Inferred Temperature Ranges for H$_2$O Ice} 
    	
         To investigate the state of H$_2$O ice on Europa's surface, we compared the JWST/NIRSpec spectra to a disk-averaged, ground-based spectrum collected with the SpeX spectrograph on NASA's Infrared Telescope Facility \citep{rayner2003IRTFSpeX} and a global Europa spectrum collected with Galileo/NIMS (Figure  \ref{H2Obands_zoomin}). We compared the 1.5 $\micron$ and 1.65 $\micron$ bands and 3.1 $\micron$ Fresnel region in all of these empirical data to synthetic spectra of crystalline and amorphous ice (Figure  \ref{H2Obands_zoomin}). 
         
         \textit{H$_2$O ice at wavelengths $<$2.5 $\micron$.} Comparison between the telescope and spacecraft reflectance spectra highlights the essentially unchanging central wavelengths of Europa's 1.5 $\micron$ and 1.65 $\micron$ band across its leading hemisphere (spectra 1-5 in Figure  \ref{H2Obands_zoomin}), which are most consistent with high concentrations of crystalline ice (model spectra a and b in Figure  \ref{H2Obands_zoomin}). The central wavelength of Europa's 1.65 $\micron$ band (1.652 -- 1.653 $\micron$) suggests a crystalline H$_2$O ice temperature range of 100 to 110 K, which falls within existing estimates of Europa's equatorial surface temperature, ranging from a nightside minimum of $\sim$80 K to a dayside maximum of $\sim$130 K \citep[e.g.,][]{spencer1999Europasurf, teolis2017Europaexomodel}. The central wavelength of Europa's 1.5 $\micron$ band (1.502 -- 1.505 $\micron$) suggests a crystalline H$_2$O ice temperature range of 90 to 150 K \citep{mastrapa2008H2Oopcon}. Because the G140H NRS1 data are saturated between 1.1 and 1.42 $\micron$, we are unable to measure the entirety of this feature, which may impact the accuracy of the 1.5 $\micron$ band's central wavelength range reported here.  
        
        The central wavelength of Europa's 2.0 $\micron$ band exhibits notable shifts, from about 1.986 $\micron$ in Tara Regio to 1.992 in other regions, and closer to 2.0 $\micron$ in disk-integrated datasets (Figure  \ref{H2Obands_zoomin}). At first glance, this range in central wavelengths (1.986 -- 2.000 $\micron$) is broadly consistent with the presence of large concentrations of amorphous ice \citep{mastrapa2008H2Oopcon}, in particular in Tara Regio (model spectra d and e in Figure  \ref{H2Obands_zoomin}). Laboratory measurements indicate that the 2.0 $\micron$ H$_2$O ice band samples shallower regolith depths ($\lessapprox$100 $\micron$) compared to the 1.5 $\micron$ and 1.65 $\micron$ bands ($\lessapprox$200 $\micron$ and $\lessapprox$500 $\micron$, respectively; \citealt{mastrapa2008H2Oopcon}), and perhaps Europa's 2.0 $\micron$ band is sampling larger concentrations of amorphous ice closer to Europa's exposed surface. However, the presence of a prominent 3.1 $\micron$ H$_2$O Fresnel peak in Tara Regio, and the ubiquitous presence of a 1.65 $\micron$ band, clearly indicates crystalline ice is exposed in this region and present at depth across Europa. Furthermore, comparison to laboratory measurements suggests the central wavelength of the 2.0 $\micron$ band in Tara Regio is best matched by cryogenic amorphous ice ($<$70 K; \citealt{mastrapa2008H2Oopcon}), which is incompatible with Europa's estimated surface temperature range (80 -- 130 K).  Instead, it seems more likely that hydrated minerals are shifting Europa's 2.0 $\micron$ band to shorter wavelengths, as demonstrated in a variety of different laboratory experiments that examined chloride salts \cite[e.g.,][]{hanley2014Europasalts, thomas2017Europabrines, deangelis2022MgCl}, especially in Tara and Powys Regiones where irradiated NaCl has been identified \citep[e.g.,][]{trumbo2022EuropaNaCl}. 
          
    	\textit{H$_2$O ice at wavelengths $>$2.5 $\micron$.} The IRTF/SpeX and Galileo/NIMS data (spectra 4 and 5 in Figure  \ref{H2Obands_zoomin}) have similar spectral properties between 3 and 4 $\micron$, with 3.6 $\micron$ continuum peaks centered near 3.67 $\micron$, albeit with minor evidence for a narrow 3.1 $\micron$ Fresnel peak in the IRTF/SpeX spectrum, but not in the Galileo/NIMS spectrum. The 3.1 $\micron$ Fresnel regions in these two spectra are most consistent with high concentrations of amorphous ice (120 K) at Europa's exposed surface (model spectra d and e in Figure  \ref{H2Obands_zoomin}), as noted in prior work \citep{hansen2004Gmoonsice}. In contrast, the three regional spectra collected with JWST/NIRSpec reveal notable spectral differences in H$_2$O ice across Europa. The northern low-latitude spectrum shows no evidence for a narrow 3.1 $\micron$ Fresnel peak, consistent with exposed amorphous ice (model spectra d and e). The 3.6 $\micron$ continuum peak in this spectrum is centered near 3.63 $\micron$, consistent with crystalline H$_2$O ice that is relatively free of contaminants \citep{filacchione2019EuropaJuno}, with an ice temperature of $\sim$120 K. The Powys spectrum shows a subtle 3.1 $\micron$ Fresnel peak, consistent with a regolith comprised of mostly amorphous H$_2$O mixed with minor amounts of crystalline ice (model spectra c and d), and a 3.6 $\micron$ continuum peak shifted to 3.69 $\micron$, consistent with warmer crystalline ice ($\sim$170 K; \citealt{filacchione2019EuropaJuno}) and/or mixing with non-ice contaminants. The Tara spectrum shows a prominent 3.1 $\micron$ Fresnel peak, consistent with high concentrations of crystalline ice mixed with a minor fraction of amorphous ice (model spectra b and c), and a 3.6 $\micron$ continuum peak shifted to 3.71 $\micron$, hinting at a crystalline ice temperature above 200 K  \citep{clark2012Iapetus} and/or substantial mixing with non-ice contaminants. 
        
        The temperatures inferred from the wavelength shift of the 3.6 $\micron$ continuum peak in spaxels spanning Powys and Tara Regiones ($>$170 K) are significantly higher than existing estimates of Europa's peak surface temperature ($\sim$130 K). Additionally, there is no evidence for thermal emission at wavelengths $>$4.5 $\micron$ that should be observed at such elevated surface temperatures, as seen in JWST/NIRSpec data of Callisto (see Figure A5 in \citealt{cartwright2024CallistoJWST}). Given the documented challenges with using the 3.6 $\micron$ continuum peak as a surface ice thermometer for regoliths that include well-mixed impurities \citep[e.g.,][]{filacchione2012SmoonsH2O}, it seems more likely that the shifted central position of this feature results primarily from the presence of contaminants mixed with H$_2$O (e.g., frozen salts, CO$_2$) as opposed to large regional variations in the surface temperature of ice. Nonetheless, if warmer ice is indeed present in Powys and Tara Regiones, it may be an indicator of recent exposure of liquid meltwater that rapidly froze once exposed at Europa's surface. 
        
        The 3.1 $\micron$ H$_2$O ice Fresnel peak samples extremely shallow depths ($<$1 $\micron$), where near-infrared photons are essentially reflected off ice grains, making it a sensitive indicator of ``pure'' H$_2$O ice (i.e., without volume scattering contributions from non-ice impurities in the underlying regolith). In Tara and Powys Regiones, the 3.1 $\micron$ Fresnel peak is centered near 3.09 $\micron$, implying a temperature between 120 to 150 K for ice grains exposed on Europa's surface \citep{mastrapa2008H2Oopcon, stephan2021H2Otempsize}. Laboratory experiments demonstrate that pure crystalline ice can shift the central wavelength of the H$_2$O ice Fresnel peak to 3.09 $\micron$ when warmed to 150 K \citep{stephan2021H2Otempsize}. Thus, the slightly shifted central wavelength of the 3.1 $\micron$ Fresnel peak in Tara and Powys ($\sim$3.09 $\micron$) may result from interactions between near-infrared photons and a thin mantle of warm crystalline ice grains, with negligible or no interactions with amorphous ice grains. 

        The intensity of the 3.6 $\micron$ continuum peak can be used to infer the prevailing size range for H$_2$O ice grains in a regolith \citep[e.g.][]{filacchione2012SmoonsH2O}, with larger, more absorbing grains ($\gtrapprox$100 $\micron$ diameters) suppressing the peak height relative to smaller grains ($\sim$1 to few 10's $\micron$ diameters; \citealt{mastrapa2009H2Oopcon}). The regional spectrum of Tara exhibits a suppressed 3.6 $\micron$ continuum peak relative to the regional spectrum of Powys, which is in turn suppressed relative to the ice-rich northern low-latitudes regional spectrum (Figure  \ref{H2Obands_zoomin}, Table  \ref{band_measurements}), suggesting that Tara's regolith may be dominated by larger grains than elsewhere on Europa's leading hemisphere. However, the lower intensity of the 3.6 $\micron$ continuum peak may also result from higher concentrations of impurities, and thus, less exposed H$_2$O ice. Supporting this interpretation, the 1.5 $\micron$, 1.65 $\micron$, and 2.0 $\micron$ bands in the regional spectrum of Tara are notably weaker than the regional spectra of Powys or Europa's northern low-latitudes (Table \ref{band_measurements}), consistent with less ice and/or a regolith dominated by smaller grains.
        
        The 3.6 $\micron$ continuum peak samples over an order of magnitude greater regolith depths ($\sim$50 $\micron$ deep) compared to the 3.1 $\micron$ Fresnel peak ($<$1 $\micron$ deep). Consequently, these two features likely probe distinct regolith layers, with the 3.1 $\micron$ Fresnel peak sensitive to a thin layer of small ice grains exposed at the surface, whereas the 3.6 $\micron$ continuum peak, and the other H$_2$O ice features examined here, sensitive to a layer (or multiple layers) of larger grains at depth. Follow-up laboratory experiments that measure reflectance spectra of samples composed of H$_2$O ice mixed with salts and other non-ice impurities are needed to better understand how grain size and temperature influence the spectral properties of Europa's 3.6 $\micron$ H$_2$O ice continuum peak. Future measurements made by the Europa Thermal Emission Imaging System (E-THEMIS; \citep{christensen2024E-THEMIS}) on Europa Clipper will provide invaluable insight on the temperature and porosity of Europa's regolith.
         
    \subsection{Spectral Properties and Spatial Distribution of CO$_2$} 
    
        Solid-state CO$_2$ is present across Europa's leading and anti-Jovian hemispheres, primarily identified by a prominent, doubled-lobed feature centered near 4.25 $\micron$ and 4.27 $\micron$ (Figure \ref{CO2_maps}). Prior analysis of Galileo/NIMS data shows that CO$_2$ is also present across Europa's trailing and sub-Jovian hemispheres \citep[e.g.][]{hansen2008EuropaCO2}, demonstrating the global presence of this molecule. The formation of the 4.25 $\micron$ lobe has been attributed to irradiation of H$_2$O in an organic-rich environment, as well as CO$_2$ deposited onto a frozen saline mixture \citep{villanueva2023EuropaJWST}. The 4.27 $\micron$ lobe could form from irradiation of a substrate composed of $^1$$^2$CO$_2$ and H$_2$O ice \citep{protopapa2024CharonJWST}.
        
\begin{figure}[h!]
         \center
			\includegraphics[scale=0.90]{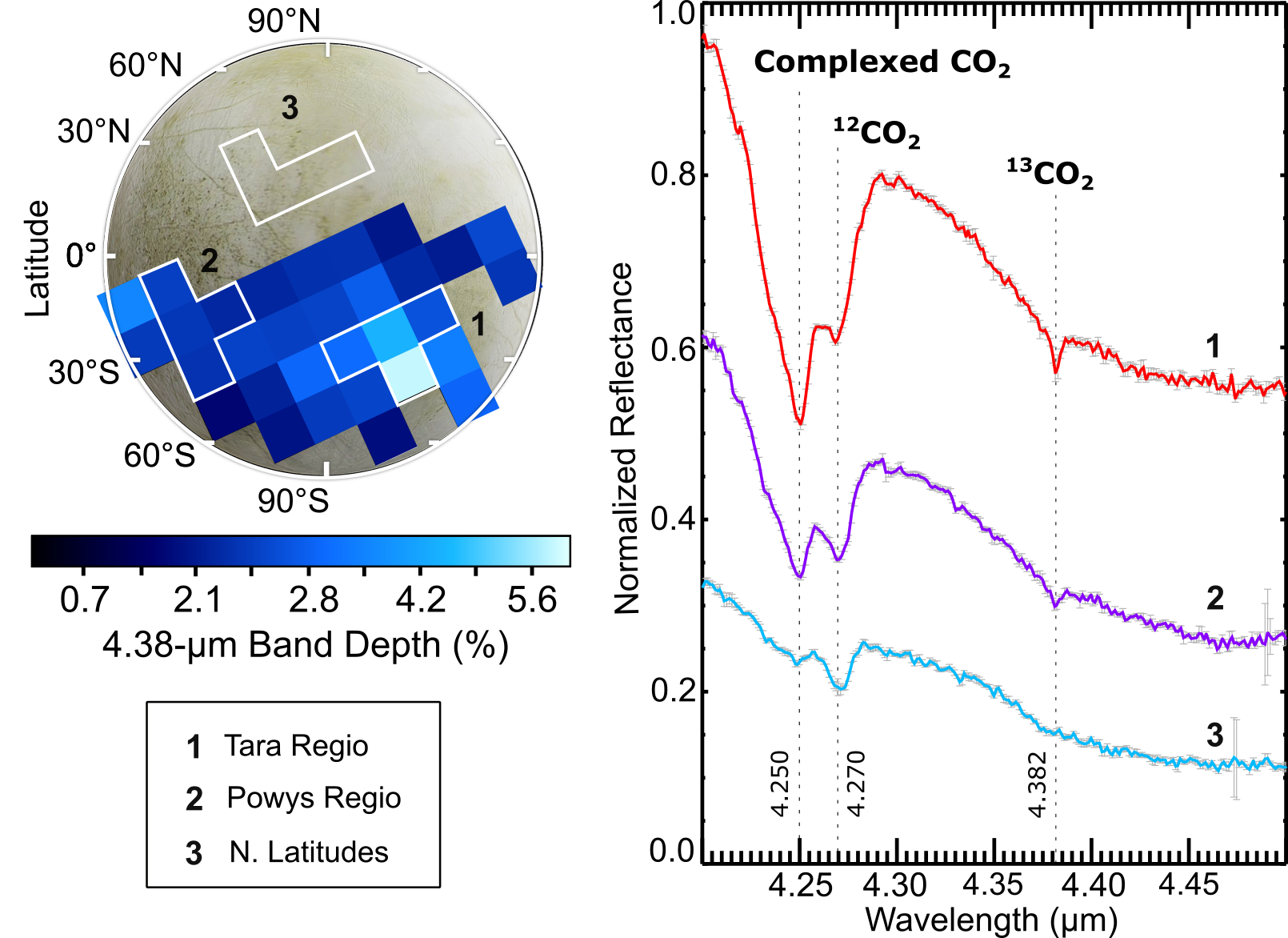}
		 \center
            \vspace{-0.1 cm}\caption{\textit{Left: Band depth measurements for the 4.38 $\micron$ band ($>$3$\sigma$ detection, also shown in Figure  \ref{CO2_maps} above). White polygons show the locations of JWST/NIRSpec spaxels incorporated into the representative regional spectra shown on the right: 1. Tara Regio (red), 2. Powys Regio (purple), 3. northern low-latitudes (blue), and their 1$\sigma$ uncertainties (gray error bars). These data are normalized to 1 at 4.0 $\micron$ and offset vertically for clarity. The central wavelengths ($\micron$) for detected solid-state CO$_2$ features are listed vertically along each dotted line.}}
        \label{CO2_compare}
    \end{figure}

        Unlike the convolved 4.25 $\micron$ and 4.27 $\micron$ CO$_2$ feature, the 4.38 $\micron$ band displays no evidence for band splitting, with a band center locked to 4.381 or 4.382 $\micron$ (Table  \ref{band_measurements} and Figures  \ref{CO2_maps} and  \ref{CO2_compare}). At first glance, the lack of band splitting suggests $^1$$^3$CO$_2$ is primarily associated with either the 4.25 $\micron$ or 4.27 $\micron$ lobe, but not both. Irradiated samples composed of CO$_2$ deposited onto frozen salts exhibit 4.25 $\micron$ lobes and $^1$$^3$CO$_2$ bands near 4.386 $\micron$ \citep{villanueva2023EuropaJWST}, representing a 0.004 to 0.005 $\micron$ wavelength shift from the center of Europa's 4.38 $\micron$ band, which should be detectable with the G395H grating (R $\sim$ 3000 near 4.4 $\micron$). Thin and thick films composed of crystalline CO$_2$ ice \citep[e.g.,][]{quirico1997near} exhibit 4.27 $\micron$ $^1$$^2$CO$_2$ and 4.38 $\micron$ $^1$$^3$CO$_2$ absorption bands consistent with the features detected on Europa, but crystalline CO$_2$ ice should sublimate rapidly at its estimated dayside surface temperatures \citep[e.g,][]{fray2009sublimation} and seems unlikely to contribute. CO$_2$ molecules isolated in a crystalline H$_2$O ice matrix (4.257 $\micron$; Eric Quirico, private communication) or CO$_2$ clathrates (4.261 and 4.277 $\micron$; \citealt{oancea2012CO2clathrates}) could persist on Europa's surface at peak dayside temperatures, albeit the central wavelengths of these CO$_2$ features are not ideal matches to Europa's 4.27 $\micron$ lobe (centered between 4.268 and 4.272 $\micron$). Alternatively, CO$_2$ trapped in amorphous H$_2$O ice ($\sim$4.273 $\micron$; e.g., \citealt{ehrenfreund1999H2OCO2}) can provide a closer match to the central wavelength of Europa's 4.27 $\micron$ lobe. CO$_2$ trapped in amorphous ice was proposed to explain Ganymede's 4.270 $\micron$ CO$_2$ feature detected at high northern latitudes on its leading hemisphere \citep{bockelee2024GanymedeJWST}. Amorphous ice is likely more prevalent on Europa than Ganymede due to its colder surface temperatures and its lack of an intrinsic magnetic field, which helps to shield ice at low latitudes on Ganymede from  amorphization via charged particle bombardment. On the other hand, peak dayside temperatures at Europa's low latitudes are more than sufficient to remove exposed amorphous ice on short timescales (see \textit{``thermal recrystallization''} subsection below), hinting that trapping of CO$_2$ in crystalline ice may be more prevalent.

        The 4.38 $\micron$ band is primarily detected over southern latitudes and is essentially absent from the northern latitudes of Europa's leading hemisphere (Figure \ref{CO2_compare}). Consequently, the 4.38 $\micron$ band is present where the 4.25 $\micron$ and 4.27 $\micron$ lobes are strongest (Figure  \ref{CO2_maps}). The central wavelength of the 4.38 $\micron$ band suggests it might be primarily spatially associated with the 4.27 $\micron$ lobe (i.e., CO$_2$ trapped in amorphous or crystalline ice). In contrast, the spatial distribution of the 4.38 $\micron$ band appears to more closely track the 4.25 $\micron$ lobe (Figure \ref{CO2_maps}), and the 4.38 $\micron$ band is only present ($>$3$\sigma$ detection) in spaxels where the band depth of the 4.25 $\micron$ lobe is greater than the band depth of the 4.27 $\micron$ lobe. Unraveling this discrepancy between the spectral properties and spatial distributions of Europa's solid-state CO$_2$ features will benefit from ongoing and future laboratory experiments that investigate the spectral properties of CO$_2$ trapped in amorphous and crystalline ice, with and without hydrated salts, under conditions relevant to Europa. Furthermore, future close flybys by Clipper/MISE will be able to more precisely map the distribution of CO$_2$ and determine whether $^1$$^3$CO$_2$ is present at northern latitudes, beyond Tara and Powys Regiones.

    \subsection{Refreshing Surface Constituents in Tara and Powys Regiones}
        The results reported here for H$_2$O ice and $^1$$^3$CO$_2$, along with prior observations that characterized H$_2$O ice \citep[e.g.,][]{ligier2016EuropaVLT}, irradiated NaCl \citep[e.g.,][]{trumbo2022EuropaNaCl}, solid-state CO$_2$ \citep{villanueva2023EuropaJWST, trumbo2023EuropaCO2}, and H$_2$O$_2$ \citep[e.g.,][]{wu2024europaH2O2,raut2024EuropaH2O2}, demonstrate that Tara and Powys Regiones are spectrally distinct from the surrounding terrains, in particular when compared to H$_2$O ice-rich regions at northern low-latitudes (Figure  \ref{H2Obands_zoomin}). The presence of a 3.1 $\micron$ Fresnel peak, diagnostic of crystalline H$_2$O ice, indicates that the exposed surfaces of Tara and Powys ($<$1 $\micron$ depths) are refreshed over timescales short enough to outpace the irradiation processes that are able to effectively amorphize ice-rich regions at northern low-latitudes and elsewhere on Europa's leading hemisphere. Refreshment of crystalline ice could be driven by thermal recrystallization of exposed ice grains outpacing amorphization in these relatively dark and warm low-latitude regions. Geologic exposure of saline meltwater in the chaos terrains that dominate Tara and Powys may help replenish crystalline ice and solid-state CO$_2$ over longer timescales.          

        \textit{Thermal recrystallization.} The surface of Europa is subjected to a substantial flux of co-rotating magnetospheric plasma and energetic charged particles, with surface bombardment patterns that are non-uniform and significantly affected by draping of the Jovian magnetic field due to Europa's atmosphere \citep[e.g.,][]{addison2021Europamagnetosphere, nordheim2022Europamagnetosphere}. The previously noted absence of a 3.1 $\micron$ H$_2$O ice Fresnel peak is generally assumed to result from intense sputtering and charged particle weathering disrupting the long-range order of crystalline ice \citep[e.g.,][]{hansen2004Gmoonsice}. However, thermal recrystallization rates for H$_2$O ice at low latitudes (30$\degree$S -- 30$\degree$N) could be competitive with amorphization rates, perhaps occurring over timescales as short as $\sim$0.15 to 2 days, for a more porous regolith (43$\%$ pore space), or $\sim$3.5 to 41 days for a more compacted regolith (9$\%$ pore space) \citep{mitchell2017Europaporosity}. Consequently, thermal recrystallization should be operating efficiently at latitudes corresponding to Tara and Powys Regiones (0 -- 30$\degree$S), where exposed crystalline ice is present, \textit{as well as} the northern low-latitude zone identified in this study (0 -- 30$\degree$N), where exposed ice is dominantly amorphous.
        
        To investigate whether thermal recrystallization may be operating over different timescales in Tara versus the northern low-latitude zone, we estimated the timescales for effective amorphization of ice by charged particle irradiation in these two locations. We utilized the calculated surface fluxes for 1 keV to 100 MeV protons, O$^n$$^+$, and S$^n$$^+$ ions \citep{nordheim2022Europamagnetosphere}, finding that the total surface energy flux (\textit{F}) from magnetospheric ions in these two locations is almost identical ($\sim$1.5*10$^8$ MeV cm$^-$$^2$ s$^-$$^1$). Using the approach described in \cite{fama2010Europaamorph}, we estimated the fraction of amorphous ice, $\phi$$_A$ = $\phi$$_A$$_m$$_a$$_x ~(1 - exp(-kFt/N))$ [Equation 1],  where the maximum fraction of amorphous ice on the surface ($\phi$$_A$$_m$$_a$$_x$) was set to 1 \citep{dalle2015Rhea}. The experimentally determined amorphization coefficient, \textit{k}, given in molecules eV$^-$$^1$ \citep{baragiola2013radiation, fama2010Europaamorph, loeffler2020crystallineH2O} was set to 0.2 molecules eV$^-$$^1$, based on experimental results for ion bombardment of H$_2$O ice at $\sim$100 K (see Figure 16.3 in \citealt{baragiola2013radiation}), a reasonable approximation of the \textit{mean} surface temperatures in these two regions. The number of H$_2$O molecules, \textit{N}, represents an icy regolith volume of 1 cm$^2$ by 10 $\micron$ deep, covering the range of depths over which protons (100's of keV to MeV) deposit most of their energy \citep[e.g.,][]{berger1999estar}. The amorphization rate, \textit{t}, is the timescale to amorphize volume \textit{N}, and represents an upper limit for this process. 
         
         By setting the fraction of amorphous ice to 63$\%$ ($\phi$$_A$ = 0.63), and solving for \textit{t} in Equation 1, we estimate that exposed crystalline ice ($\leq$10 $\micron$ depths) should be amorphized by magnetospheric ion bombardment on timescales of $\sim$15 days at low latitudes on Europa's leading hemisphere. Because Tara and the northern low-latitude zone receive similar input from magnetospheric ion bombardment, our calculation indicates that some other important difference exists between the properties of the exposed regoliths in these two locations. We suggest that Tara, and by extension Powys, are mantled by porous layers of ice grains that rapidly recrystallize, thereby outpacing amorphization over timescales $<$15 days, which is more consistent with the $\sim$0.15 to 2 days recrystallization estimates for the high porosity case presented in \cite{mitchell2017Europaporosity}. In contrast, the northern low-latitude zone is mantled by a more compacted layer of ice grains, where recrystallization operates over timescales $>$15 days and amorphization processes dominate, which is more consistent with the $\sim$3.5 to 41 days recrystallization estimates for the low porosity case presented in \cite{mitchell2017Europaporosity}. 
         
         Supporting this interpretation, Tara and Powys Regiones exhibit lower bond albedos (\textit{B} = 0.4 -- 0.6) than the northern low-latitude zone (\textit{B} = 0.7 -- 0.8; \citealt{mergny2025Europaalbmap}), consistent with warmer temperatures and faster rates of recrystallization in these regiones. Tara and Powys also have lower estimated thermal inertias and higher brightness temperatures ($\sim$100 K) compared to the northern low-latitude zone ($\sim$80 K), based on the analysis of Band 7 (0.87 mm) Atacama Large Millimeter/submillimeter Array (ALMA) observations, which probe $\sim$10 to 20 cm depths \citep{cordiner2024EuropaALMA}. Furthermore, analysis of Band 3 (3.05 mm) ALMA data suggests the porosity of regolith material on Europa's leading hemisphere ranges between 40 to 70$\%$ at depths of $\sim$500 to 1000 $\micron$, with Tara and Powys likely being more porous than the ice-rich, northern low-latitude zone \citep{thelen2024Europa_ALMA}. A different study that analyzed ALMA data \citep{trumbo2018EuropaALMA} also derived higher thermal inertia, or lower emissivity, estimates for an anomalous region co-located with the northern low-latitude zone, suggesting larger particle sizes or higher transparency to sunlight in this region \citep{trumbo2018EuropaALMA}.          
        
        Thus, prior studies support our interpretation that a colder, brighter, and lower porosity regolith, possibly dominated by larger ice grains, mantles Europa's northern low-latitude zone, where amorphization outpaces recrystallization. Similarly, these studies support our interpretation that a warmer, darker, and higher porosity regolith, possibly dominated by smaller ice grains, mantles Tara and Powys Regiones, where recrystallization outpaces amorphization.  Below, we consider a few different processes that could help explain these disparate regolith properties and the prevalence of exposed crystalline ice and solid-state CO$_2$ in Tara and Powys Regiones. 
        
        \textit{Exposed during chaos formation?} Fresh crystalline ice and hydrated salts could result from the exposure and subsequent freezing of meltwater due to chaos formation. Of the currently hypothesized chaos formation mechanisms (see \citealt{daubar2024Clippergeol}), our results are more consistent with the rapid freezing of saline meltwater, perhaps resulting from collapsing and refreezing of liquid H$_2$O lenses  \citep[e.g.,][]{collins2000Europadiapir, schmidt2011Europachaos, michaut2014EuropaH2Osills}. This chaos formation mechanism is favored by detailed morphological studies of chaos terrains \citep{leonard2018Europachaos}, although other mechanisms involving exposure of saline liquid H$_2$O could hypothetically reproduce the spectral results reported here. In our preferred scenario, saline liquid H$_2$O containing dissolved CO$_2$ or other carbon-bearing compounds resides in melt-pockets. This saline liquid mixture begins to freeze and expand in the ice shell, eventually causing the mixture to breach Europa's surface where it then rapidly freezes, locking in entrained CO$_2$ molecules, or organic content that is radiolytically-modified into CO$_2$ post-emplacement. This mixture could explain the presence of Europa's 4.25 $\micron$ lobe, which results from CO$_2$ complexed with salts \citep{villanueva2023EuropaJWST}. Molecular interactions between CO$_2$ and refractory salts on Europa are presumably consistent with prior detection of a prominent 4.257 $\micron$ feature by Galileo/NIMS that was attributed to CO$_2$ complexed with refractory dark material \citep[e.g.,][]{carlson1996GmoonsNIMS, mccord1997NIMS, mccord1998NIMS, hibbitts2000distributions}. In this scenario, a thin layer of H$_2$O frost might also condense, possibly capturing CO$_2$ molecules in the rapidly forming crystalline structure. Alternatively, trapped organics are irradiated, reacting with surrounding H$_2$O molecules to form $^1$$^2$CO$_2$ and $^1$$^3$CO$_2$, thereby explaining the origin of the 4.27 $\micron$ lobe and the 4.38 $\micron$ band, respectively. Thus, chaos formation mechanisms that result in exposure of internally-derived saline meltwater might be an important source of crystalline H$_2$O ice and CO$_2$ in Tara and Powys Regiones, albeit subsequent amorphization and recrystallization cycles would quickly obscure the spectral signature of internally-derived ice.
        
        \textit{Plume deposits?} Crystalline ice in Tara and Powys may result from sporadic, CO$_2$ gas-driven south polar plume activity \citep{roth2014Europaplume}, or hitherto undetected localized geysers in Tara Regio. In either scenario, ice grains are rained onto Europa's surface, primarily over its southern latitudes. Plume activity could be complemented by lower level outgassing of H$_2$O vapor that condenses once exposed at the surface. Plume grains that rain onto Europa's surface could be initially crystalline, or they could be crystallized by subsolar heating, especially if they land on darker and warmer surfaces, such as Tara and Powys Regiones. Deposition of plume material might explain the stronger 3.1 $\micron$ Fresnel region at southern latitudes reported here and previously observed by Juno/JIRAM \citep{filacchione2019EuropaJuno}. However, it is uncertain whether a south polar plume would be sufficiently energetic to mantle the entirety of Europa's southern latitudes with ice grains, which is required to explain the spatial distribution of crystalline ice reported here.
        
        Whether this mechanism can explain the spatial association between Tara and Powys and solid-state CO$_2$ is also uncertain. Although outgassed CO$_2$ molecules could hypothetically co-condense with H$_2$O on chaos terrains, it is doubtful that CO$_2$ could persist at Europa's peak dayside surface temperatures ($\sim$130 K). Deposition of plume material that includes CO$_2$ complexed with frozen salts, or organic material subsequently irradiated and oxidized to CO$_2$, should be more stable on Europa's surface and could account for Europa's 4.25 $\micron$ lobe. However, if complexed CO$_2$ primarily originates from plume fallout, it is uncertain why it is concentrated in Tara, unless sourced by undetected localized geysers.         

        \textit{Volatile migration and condensation?} Perhaps sublimated H$_2$O vapor and CO$_2$ gas, migrating from Europa's warmer and darker trailing hemisphere, condense on the nightside leading hemisphere, contributing to a transient frost layer. Subsequent dayside heating then preferentially recrystallizes condensed H$_2$O frost on the darker and warmer surfaces of Tara and Powys Regiones compared to the brighter and colder, ice-rich northern low-latitudes. This scenario cannot readily account for the concentration of CO$_2$ in Tara and Powys, as this volatile should display a strong preference for condensation at high latitudes and in colder and brighter icy terrains, counter to the observed spatial distribution, and an internal origin for CO$_2$ therefore seems more likely.   
                       
        \textit{Impact exposure?} Collisions with micrometeorite grains and large-scale impactors overturn regolith materials on icy moons over geologic timescales, exposing fresh crystalline H$_2$O ice, CO$_2$, and other compounds that have been shielded at depth from irradiation \citep[e.g.,][]{zahnle2003cratering, costello2021Europaimpactgarden}. On tidally-locked objects like Europa, impact gardening is hypothesized to operate more efficiently on their leading hemispheres, especially near their apexes (0$\degree$ latitude, 90$\degree$W). Our results show that exposed crystalline H$_2$O ice and CO$_2$ are not concentrated at Europa's apex, and instead are found at southern low-latitudes, near the center of Tara Regio ($\sim$10 -- 30$\degree$S and 75$\degree$W). In contrast, the northern low-latitude zone ($\sim$10 -- 30$\degree$N and 105$\degree$W) shows negligible evidence for exposed crystalline ice or $^1$$^3$CO$_2$ (spectrum 3 in Figure \ref{H2Obands_zoomin}). If impact gardening is the primary driver of regolith refreshment in Tara, then some other, currently unknown, mechanism is needed to explain why dust grains preferentially strike Europa's southern hemisphere and do not equally refresh H$_2$O ice at northern low-latitudes. Furthermore, micrometeorite exposure of fresh crystalline ice likely operates at considerably slower rates than the $\sim$15 day estimate for amorphization reported here, and it seems unlikely that impact gardening would play an equally prominent role in shaping Europa's exposed regolith. A large impact that exposed fresh ice or meltwater in the very recent past may explain the presence of crystalline H$_2$O ice in Tara. However, the geographic scale of the 3.1 $\micron$ Fresnel peak detected by JWST/NIRSpec (Figure \ref{H2O_maps2}) suggests that the resulting impact crater would be spatially extensive and presumably visible in Galileo images ($\gtrapprox$1 km/pixel),  but there is no evidence for an impact feature or a large ejecta blanket in Tara or Powys. 
        
	\section{Summary and Conclusions} 
				
	    We analyzed JWST/NIRSpec observations of Europa ($\sim$1.48 -- 5.35 $\micron$) to investigate the origin of crystalline H$_2$O ice and solid-state CO$_2$. For H$_2$O ice, we made continuum-divided measurements of Europa's 1.5 $\micron$, 1.65 $\micron$, and 2.0 $\micron$ H$_2$O absorption bands (Figure  \ref{H2O_maps1}) as well as of its 3.1 $\micron$ H$_2$O Fresnel peak and its 3.6 $\micron$ H$_2$O continuum peak (Figure  \ref{H2O_maps2}). For CO$_2$, we made band measurements of the 4.25 $\micron$ and 4.27 $\micron$ lobes of Europa's strongest solid-state CO$_2$ feature and its 4.38 $\micron$ $^1$$^3$CO$_2$ absorption band (Figure \ref{CO2_maps}). We also generated and compared the properties of three regional spectra, representative of Tara Regio, Powys Regio, and Europa's northern low-latitudes (Figures  \ref{example_spectra}, \ref{H2Obands_zoomin}). The resulting measurements and spectral maps for the 1.65 $\micron$ band demonstrate that crystalline H$_2$O ice is present beneath the exposed surface ($\gtrapprox$300 $\micron$ depths) across Europa, with the strongest signature at northern low-latitudes, consistent with prior observations \citep[e.g.,][]{hansen2004Gmoonsice, ligier2016EuropaVLT}. Our measurements of the 3.1 $\micron$ H$_2$O ice Fresnel peak demonstrate that crystalline H$_2$O ice is largely absent from the exposed surface of Europa's northern hemisphere ($<$1 $\micron$ depths), but it is present at the exposed surface of its southern hemisphere, especially in Tara and Powys Regiones. The disparity between these measurements supports the hypothesis that Europa's regolith is vertically stratified \citep{hansen2004Gmoonsice}, with crystalline ice dominating at depth and amorphous ice dominating at the exposed surface of Europa's leading hemisphere, except in Tara and Powys Regiones, where crystalline ice dominates both at depth and at the exposed surface (Figures \ref{H2Obands_zoomin} and \ref{H2O_maps2}). 
        
        We estimate that charged particle-driven amorphization effectively removes the signature of exposed crystalline ice from low latitudes on Europa's leading hemisphere in $\sim$15 days, and the presence of crystalline ice at the topmost surface of Tara and Powys Regiones indicates ongoing refreshment over shorter timescales (perhaps $\ll$15 days). Because the 3.1 $\micron$ Fresnel peak is sensitive to extremely shallow depths, it only samples exposed ice grains that may form a porous H$_2$O frost layer, mantling the relatively darker and warmer surfaces of Tara and Powys Regiones. Thermal recrystallization of amorphous ice contained within such a porous layer should outpace amorphization rates, in particular for low latitude regions near the subsolar point. Our results therefore suggest that the 3.1 $\micron$ crystalline H$_2$O ice Fresnel peak is a spectral tracer of recent (and presumably ongoing) surface modification. 
        
        The origin of such a porous H$_2$O frost layer is uncertain, and it could arise from a variety of processes including exposure of saline meltwater, sporadic plume activity, outgassing of H$_2$O vapor, sublimation and migration of H$_2$O vapor from Europa's dayside trailing hemisphere, or recent regolith overturn by impacts. Exposure of saline meltwater in chaos terrains may be the most likely process given the presence of other, likely internally-derived species in Tara and Powys Regiones, such as NaCl and CO$_2$. Indeed, our analysis of Europa's double-lobed CO$_2$ feature and its 4.38 $\micron$ $^1$$^3$CO$_2$ band favors an endogenic origin for CO$_2$, consistent with prior work \citep{villanueva2023EuropaJWST, trumbo2023EuropaCO2}. However, the physical state of CO$_2$ on Europa's surface is less certain, and it is likely complexed with more refractory components, such as frozen salts, and trapped in crystalline or amorphous H$_2$O ice. Finally, the results reported here provide high signal-to-noise context for the upcoming Europa Clipper mission that will make many close passes of Europa, collecting high spatial resolution footprints with its NIR spectrometer, MISE (1 -- 5 $\micron$).

	\section{Acknowledgments} 

        This work is based [in part] on observations made with the NASA/ESA/CSA James Webb Space Telescope. The data were obtained from the Mikulski Archive for Space Telescopes at the Space Telescope Science Institute, which is operated by the Association of Universities for Research in Astronomy, Inc., under NASA contract NAS 5-03127 for JWST. These observations are associated with GTO program 1250. Support for the analysis of program 1250 data was provided in part by MISE subcontract $\#$ 1715089.

	\bibliography{References_spring2025.bib}{}
		\bibliographystyle{aasjournal}
		
		
		
		\renewcommand{\thesubsection}{A\arabic{subsection}}
		\setcounter{subsection}{0}
		
		\renewcommand{\thefigure}{A\arabic{figure}}
		\setcounter{figure}{0}
		
		\renewcommand{\thetable}{A\arabic{table}}
		\setcounter{table}{0}	

\end{document}